\documentclass[12pt]{iopart}   

\usepackage{iopams}
\usepackage{braket}
\usepackage[english]{babel}
\usepackage{graphicx}
\usepackage{dcolumn}
\usepackage{bm}
\usepackage{verbatim}
\usepackage{amsfonts}
\usepackage{amssymb}
\usepackage{color,soul}
\sethlcolor{yellow}
\def\be#1{\begin{equation}\label{#1}}
\def\ee{\end{equation}}
\def\bea#1{\begin{eqnarray}\label{#1}}
\def\eea{\end{eqnarray}}
\def\Bea{\begin{eqnarray*}}
\def\Eea{\end{eqnarray*}}
\def\sp{\hspace{.5em}}
\def\sph{\hspace{.25em}}

\def\Eq#1{\eref{#1}}   

\def\no{\nonumber \\}

\def\tit#1{\textit{#1}}

\def\displ{\displaystyle}

\def\mbf#1{\mbox{{\boldmath $#1$}}}
\def\half{\frac{1}{2}}
\def\ket#1{|#1\rangle}
\def\bra#1{\langle #1|}
\def\a{\alpha}
\def\b{\beta}

\def\d{\delta}

\def\l{\lambda}
\def\L{\Lambda}
\def\u{\mu}
\def\v{\nu}

\def\s{\sigma}
\def\t{\tau}
\def\pr{\prime}
\def\dag{\dagger}
\def\bvec#1{\mbf{#1}}
\def\hbvec#1{\hat{\mbf{#1}}}

\def\eps{\epsilon}
\def\om{\omega}
\def\Om{\Omega}

\def\M{{\mathcal{M}}}
\def\U{{\mathcal{U}}}
\begin{document}

\title{Observer dependent entanglement }

\author{Paul M. Alsing}
\address{Air Force Research Laboratory, Information Directorate, Rome, N.Y., USA}
\author{Ivette Fuentes}
\address{School of Mathematical Sciences, University of Nottingham, Nottingham NG7 2RD United Kingdom}
\begin{abstract}
Understanding the observer-dependent nature of quantum entanglement has been a central question in relativistic quantum information. In this paper we will review key results on relativistic entanglement in flat and curved spacetime and discuss recent work which shows that motion and gravity have observable effects on entanglement between localized systems.

\end{abstract}


\section{Introduction}\label{Introduction}
In quantum information non-classical properties such as entanglement are exploited to improve information tasks. A prototypical example of this is quantum teleportation where two observers Alice and Bob use two quantum systems in an entangled state to transmit information about the state of a third system.  Impressively, cutting-edge experiments involving entanglement based communications are reaching regimes where relativistic effects can no longer be neglected.  Such is the case of protocols which involve distributing entanglement over hundreds of kilometers \cite{PhysRevLett.98.010504,Zeilinger2007}.
Understanding entanglement in relativistic settings has been a key question in relativistic quantum information.  Early results show that entanglement is observer-dependent \cite{Alice,PMAlsing:IFuentesSchuller:RBMann:TETessier:06,
AlsingMilburn03,PMAlsing:DMcMahon:GJMilburn:08}.  The entanglement between two field modes is degraded by the Unruh effect when observers are in uniform acceleration.  We also learned that the spatial degrees of freedom of global fields are entangled, including the vacuum state \cite{Resnik:03,PhysRevA.71.042104,SYLin:BLHu:10,Olson:2011bq}. This entanglement can be extracted by point-like systems and in principal be used for quantum information processing (see for example \cite{Lin2012a,cqedsabin,time-liketeleportation,PhysRevLett.109.033602}).  Most of the early studies on relativistic entanglement in non-inertial frames involved  global modes.  However,  more recently, researchers in the field  have focused their attention on understanding entanglement between fields or systems which are localized in space and time. The motivation for this is that entangled localized systems can be in principle measured, transformed and exploited for quantum information tasks. Among the most popular systems considered for this purpose are moving cavities \cite{TGDownes:IFUentes:TCRalph:10,DEBruschi:JLouko:IFuentes:12,NFriis:DEBruschi:JLouko:IFuentes:12,NFriis:ARLee:DEBruschi:JLouko:10,DEBruschi:Dragan:Lee:JLouko:IFuentes:12,NFriis:IFuentes,Friss:Hubber:Bruschi:IFuentes:12}, point-like detectors \cite{Lin2012a,Lin2008s,Lin2009,Lin2008v,PhysRevA.81.062320} and localized wave-packets \cite{QCommRalph,dragan,dragan2}. In this paper we will review global mode entanglement in flat and curved spacetime which constitutes the first step in the study of entanglement in quantum field theory. We will then discuss more recent ideas on entanglement which show that motion and gravity have observable effects on quantum correlations between localized systems  \cite{DEBruschi:JLouko:IFuentes:12,NFriis:DEBruschi:JLouko:IFuentes:12,NFriis:ARLee:DEBruschi:JLouko:10,DEBruschi:Dragan:Lee:JLouko:IFuentes:12,NFriis:IFuentes,Friss:Hubber:Bruschi:IFuentes:12}.
Interestingly, in these settings it is possible to generate quantum gates through motion in spacetime \cite{DEBruschi:Dragan:Lee:JLouko:IFuentes:12,Friss:Hubber:Bruschi:IFuentes:12,newjorma}.

The observer-dependent nature of entanglement is a consequence of the particle content being different for different observers in quantum field theory \cite{B&D,JL:Ball:IFuentesSchuller:FPSchuller:06}. In flat spacetime, all inertial observers agree on particle number and therefore, on entanglement. Entanglement is well defined in that case since inertial observers play a special role. However, in the case of curved spacetime, the entanglement in a given state varies even for inertial observers (see discussion in \cite{JL:Ball:IFuentesSchuller:FPSchuller:06}).

In special relativity one also finds that quantum correlations are observer-dependent. The entanglement between two spin particles is invariant only when the spin and momentum of the particles are considered to be a single subsystem. If only spin degrees of freedom are considered, different inertial observers would disagree on the amount of the entanglement between the particles. Some works show that spin entanglement in transformed into momentum entanglement  under Lorentz transformations while
some
recent papers argue that considering spin degrees of freedom alone (by tracing over momentum) lead to inconsistencies (this will be discussed further in section \ref{LTs}).

The paper is organized as follows:
in the section (\ref{Technical_tools}) we will introduce technical tools in quantum field theory and quantum information.  We will review the basics of field quantization focusing on the free bosonic massless case. We will describe the interaction of the field with point-like systems better known as Unruh-DeWitt detectors.  By imposing boundary conditions we will describe fields contained within moving mirrors (cavities) and show how to construct wave-packets that are localized in space and time. A brief discussion on fields in curved spacetimes will be presented. We will end the section by reviewing measures of entanglement in the pure and mixed case as well as introduce the covariant matrix formalisms which allows for relatively simple entanglement computations in quantum field theory.
In section (\ref{Entanglement_of_global_modes}) of this paper we will review the results on free mode entanglement  in non-inertial frames, in an expanding universe  and in a black hole spacetime. We will present ideas on how to extract field entanglement using Unruh-DeWitt detectors in section  (\ref{Extracting_global_mode_entanglement}).
In section (\ref{Accelerated_cavities}) we will present a more modern view on the study of entanglement in quantum field theory  where the entanglement between the modes of moving cavities is analyzed and review recent work on how localized wave-packets can be used to implement quantum information protocols.
For completeness, in section (\ref{LTs}) we review the concept of observer dependent entanglement for the case of zero acceleration. Here we discuss the Wigner rotation, the change in state under Lorentz transformations and their effect on entanglement for spin $\half$ particles and photons.
Finally, in section (\ref{Open_questions}) we will point out open questions, discuss work in progress and future directions in the understanding of entanglement in quantum field theory.

\section{Technical tools}\label{Technical_tools}

\subsection{Quantum field theory}\label{QFT}

	The theoretical framework in which questions of relativistic entanglement are analyzed is quantum field theory in flat and curved spacetime. In the absence of a consistent quantum theory of gravity, quantum field theory allows the exploration of some aspects of the overlap of relativity and quantum theory by considering quantum fields on a classical spacetime. The most important lesson we have learned from quantum field theory is that fields are fundamental, while particles are derived notions (if at all possible) \cite{B&D}.  Field quantization is inequivalent for different observers and therefore, the particle content of the field may vary for different observers. For example, the Minkowski vacuum seen by inertial observers in flat spacetime corresponds to a state populated with a thermal distribution of particles for observers in uniform acceleration \cite{WGUnruh:76}.  The temperature, known as the Unruh temperature, is a function of the observer's acceleration. As we will see, a consequence of this is that the entanglement of free field modes in flat spacetime is observer-dependent \cite{Alice,PMAlsing:IFuentesSchuller:RBMann:TETessier:06},
and effects quantum information processing tasks such a teleporation \cite{AlsingMilburn03,AlsingMcMahonMilburn04,Lin2012a}.
Another interesting example is that of an expanding universe \cite{B&D}. The vacuum state for observers in the asymptotic past is populated by particles as seen by observers in the future infinity \cite{B&D}. The expansion of the universe creates particles and these particles are entangled \cite{JL:Ball:IFuentesSchuller:FPSchuller:06,PhysRevD.82.045030}.  It might at first sight seam surprising that the dynamical Casimir effect is closely related to the Unruh effect \cite{EYablonovitch:93,VVDodonov:10}.  Both effects are predictions of quantum field theory. The vacuum state of an inertial cavity defined by inertial observers is inequivalent to the vacuum state of the cavity undergoing uniform acceleration as seen by observers moving along with the cavity (Rindler observers)\cite{TGDownes:IFUentes:TCRalph:10,DEBruschi:JLouko:IFuentes:12,NFriis:DEBruschi:JLouko:IFuentes:12,NFriis:ARLee:DEBruschi:JLouko:10,DEBruschi:Dragan:Lee:JLouko:IFuentes:12,NFriis:IFuentes,Friss:Hubber:Bruschi:IFuentes:12}. Therefore, if a cavity is at rest and the field is in the vacuum state, entangled particles will be created when the cavity subsequently undergoes non-uniform accelerated motion \cite{DEBruschi:JLouko:IFuentes:12,NFriis:ARLee:DEBruschi:JLouko:10}.  Related to this effect is the dynamical Casimir effect where the mirrors of the cavity oscillate \cite{VVDodonov:10,ALambrecht:MTJaekel:SReynaud:96}.  Before we discuss in more detail the entanglement between the modes of a quantum field in these and other scenarios we will revisit basic concepts in quantum field theory, considering the simplest case: the massless uncharged bosonic field (which we denote $\phi$ ) in a flat (1+1)-dimensional spacetime.  Throughout our paper we will work in natural units $c=\hbar=1$ and the signature of the metric $(+,-)$.

\subsubsection{Global fields}\label{Global_fields}
The massless real bosonic quantum field obeys the Klein-Gordon equation $\square\phi=0$,
 where the d'Alambertian operator $\square$ is defined as
\be{QFT1}
\square\phi:=\frac{1}{\sqrt{-g}}\partial_{\mu}(\sqrt{-g}g^{\mu\nu}\partial_{\nu}\phi),
\ee
where $g=det(g_{ab})$ and $\partial_{\mu}=\frac{\partial}{\partial x^{\mu}}$.
In flat $(1+1)$-dimensional spacetime the metric is $g_{\mu\nu}=\eta_{\mu\nu}=\{+-\}$ and thus,
$\square\phi=\partial^2_t-\partial^2_x$.
Minkowski coordinates $(t,x)$ are a convenient choice for inertial observers.
The solutions to the equation are plane waves
\begin{equation}\label{eq:planem}
u_{\omega,M}(t,x)=\frac{1}{\sqrt{4\pi\omega}}e^{-i\omega(t-\epsilon x)},
\end{equation}
where the label $M$ stands for Minkowski and $\epsilon$ takes the value $+1$ for modes with positive momentum (right movers) and $-1$ for modes with negative momentum (left movers). The modes of frequency $\omega>0$  are orthonormal with respect to the Lorentz invariant inner product
\be{IP}
(\phi,\psi)=-i\int_{\Sigma}(\psi^*\partial_{\mu}\phi-(\partial_{\mu}\psi^*)\phi)d\Sigma^{\mu},
\ee
where $\Sigma$ is a spacelike hypersurface.  These solutions are known as \textit{global} field modes.

To quantize the field the notion of a time-like Killing vector field is required.  A Killing vector field  $K^{\mu}$ is the tangent field to a flow induced by a transformation which leaves the metric invariant.  This means that the Lie derivative of the metric tensor defined by
\[\mathcal{L}_Kg_{\u\v}=K^\l \partial_\l g_{\u\v} + g_{\u\l}\partial_\v K^\l  +g_{\v\l}\partial_\u K^\l,\]
must vanish.
When a spacetime admits such a structure it is possible to find a special basis for the solutions of $\square\phi=0$ such that
\begin{eqnarray*}
\mathcal{L}_Ku_{k,M}&=&K^{\mu}\partial_{\mu}u_{k,M}=-i\omega u_{k,M},
\end{eqnarray*}
where we have considered the action of a Lie derivative on a function.  Vectors lying within the light cone at each point are called time-like. Therefore, if $K^{\mu}$ is a timelike Minkowski vector field, the Lie derivative corresponds to $\partial_t$. By the action of the Lie derivative on the solutions of the Klein-Gordon equation we can identify the parameter $\omega>0$ with a frequency,
and classify the plane waves such that $u_{k,M}$ are positive frequency solutions and
$u_{k,M}^*$ are negative frequency solutions.
A few words about the physical significance of the existence of a Killing vector field are in order. If a spacetime has
as Killing vector $K^\mu$, one can always find a coordinate system in which the metric is independent of one of the
coordinates and the quantity $E=p_\mu K^\mu$ is constant along a geodesic with tangent vector $p^\mu$ \cite{carroll}.
The quantity $E$ can be considered as the \textit{conserved energy} of a photon with 4-momentum $p^\mu$.
For static observers, i.e. those whose 4-velocity $U^\mu=dx^\mu/d\tau$ is proportional to the timelike Killing vector
$K^\mu$ as $K^\mu = V(x) U^\mu$, one defines the ``redshift'' factor $V = (K^\mu K_\mu)^{1/2}$ as the norm of the Killing vector (since $U^\mu U_\mu=1$). The frequency $\omega$ of the photon measured by a static observer with 4-velocity $U^\mu$ is given by $\omega = p_\mu U^\mu$, and hence $\omega = E/V$.
A photon emitted by a static observer $1$ will be observed by a static observer $2$ to have
frequency $\omega_2 = \omega_1 V_1/V_2$. Note that along the orbit of the Killing vector  $K^\mu$ (not necessarily a geodesic), $V$ is constant.
For a general $1+1$ spacetime with coordinates $x=(x^0,x^1)$, a photon $p^\mu = (\omega_0, \pm k(x))$ of frequency $\omega_0 > 0$ and
wavevector of magnitude $k(x) = \omega_0\,\sqrt{-g_{00}(x)/g_{1}(x)}$ (such that $g_{\mu\nu} p^\mu p^\nu=0$) will be measured to
have frequency  $\omega_K(x) = \omega_0\,\sqrt{g_{00}(x)}\,\sqrt{(1\pm\alpha)/(1\mp\alpha)}$ with $x$-dependent
Doppler factor $\alpha = \sqrt{-g_{00}(x)/g_{1}(x)}\,\big(K^1(x)/K^0(x)\big)$
by a static observer along the orbit of the Killing vector $K = K^0(x)\partial_{x^0} + K^1(x)\partial_{x^1}$.
In particular, in flat Minkowski spacetime with metric $g_{\mu\nu} = (+,-)$ in $(t,x)$ coordinates
a photon of frequency $\omega_0$ emitted by an inertial Minkowski observer will be measured to have the frequency $\omega_K(x) = \omega\,\,\sqrt{(1\pm \alpha)/(1\mp\alpha)}$ with Doppler factor
$\alpha = K^1(x)/K^0(x)$.


If the metric is \textit{static} ($\partial_0 g_{\mu\nu}=0$ and $g_{0\nu}=0$) then the metric components are independent of the
time coordinates $t$ and the Klein-Gordon equation can be separated into space and time components as
$f_\omega(t,\vec{x}) = e^{-i\omega t} \bar{f}_\omega(\vec{x})$ (here $(t,x)$ are general $1+1$ coordinates). The modes $(f_\omega,f^*_\omega)$ form a basis of the wave equation from which to define the notion of particles. By definition, a detector measures the proper time $\tau$ along its trajectory. If the detectors's trajectory follows the orbit of the Killing field (i.e. the static observers defined above)
the proper time will be proportional to the Killing time $t$. Modes that are positive frequency with respect to this Killing vector serve as a natural basis for describing the Fock space of particles \cite{carroll}.
Most importantly, under Lorentz transformations, timelike vectors are transformed into timelike vectors, so that the separation of modes into positive and negative frequencies remains invariant under boosts.
In a general curved spacetime, the non-existence of a Killing field implies that the separation of modes into positive and negative frequencies is different along each point of the detectors's trajectory, and hence the concept of ``particle" is lost (for further details, see \cite{B&D} and Chap. 9 of \cite{carroll}). Note that the photons of measured frequency $\omega_K(x)$ in the previous paragraph are not pure plane waves along the Killing orbit, and therefore must be decomposed into the natural positive and negative frequency modes $(f_\omega,f^*_\omega)$.

Having identified positive and negative modes, the quantized field satisfying $\square\hat\phi=0$ is then given by the following operator value function \[\hat{\phi}=\int(u_{k,M}a_{k,M}+u_{k,M}^*a_{k,M}^{\dag})dk,\] where the creation and annihilation Minkowski operators $a_{k,M}^{\dagger}$ and $a_{k,M}$  satisfy the commutation relations $[a^{\dagger}_{k,M},a_{k^{\prime},M}]=\delta_{k,k^{\prime}}$.  Note that the solutions have been treated differently by associating creation and annihilation operators with negative and positive frequency modes, respectively. The vacuum state is defined by the equation  $a_{k,M}{|0\rangle}^{\mathcal{M}}=0$ and can be written as ${|0\rangle}^{\mathcal{M}}=\prod_k {|0_{k}\rangle}^{\mathcal{M}}$ where ${|0_{k}\rangle}^{\mathcal{M}}$ is the ground state of mode $k$. Particle states are constructed by the action of creation operators on the vacuum state
\[\ket{n_1,...,n_{k}}^{\mathcal{M}}= (n_1!,...,n_{k}!)^{-1/2} \, a_{1,M}^{\dag n_1}...a_{k,M}^{\dag n_k}\ket{0}^{\mathcal{M}}.\]
Only when there exists a time-like Killing vector field it is meaningful to define particles. Observers flowing along timelike Killing vector fields are those who can properly describe particle states. This has important consequences to relativistic quantum information since  the notion of particles (and therefore, subsystems) are indispensable to store information and thus, to define entanglement. However, in the most general case, curved spacetimes do not admit time-like Killing vector fields.

Interestingly, in the case where the spacetime admits a global timelike Killing vector field, the vector field is not necessarily unique. Consider two time-like Killing vector fields $\partial_ T$ and $\partial_{\hat{T}}$. It is then possible to find in each case a basis for  the solutions to the Klein-Gordon equation   $\{{u}_k,{u}_k^*\}$ and $\{\bar{u}_k,\bar{u}_k^*\}$ such that classification into positive and frequency solutions is possible with respect to $\partial_T$ and $\partial_{\hat{T}}$ respectively.  The field  is equivalently quantized in both bases, therefore
  \[\hat{\phi}=\int(u_ka_k+u_k^*a_k^{\dag})dk=\int(\bar{u}_{k^{\prime}}\bar{a}_{k^{\prime}}+\bar{u}_{k^{\prime}}^*\bar{a}_{k^{\prime}}^{\dag})dk^{\prime}.\]
Using the inner product,  one obtains a transformation between the mode solutions and
correspondingly, between the creation and annihilation operators,
\[a_k=\sum_{k^{\prime}}(\alpha^{\ast}_{kk^{\prime}}\bar{a}_{k^{\prime}}-\beta^{\ast}_{kk^{\prime}}\bar{a}_{k^{\prime}}^{\dag}),\]
where $\alpha_{kk^{\prime}}=(u_{k},\bar{u}_{k^{\prime}})$ and  $\beta_{kk^{\prime}}=-(u_{k},\bar{u}^{\ast}_{k^{\prime}})$ are called Bogoliubov coefficients.
Since the vacua are given by
${a}_k{\ket{0}}={\bar{a}}_k{\bar{\ket{0}}}=0$
it is possible to find a transformation between the states in the two bases. We note that as long as one of the Bogoliubov coefficients $\beta_{kk^{\prime}}$ is non-zero, and the un-barred state is the vacuum state, the state in the bared basis is populated with particles. Therefore, different Killing observers observe a different particle content in the field, i.e. particles are observer-dependent notions.

In flat spacetime there are two kinds of observers who can meaningfully describe particles for all times: inertial observers who follow straight lines and observers in uniform acceleration who's trajectories are given by hyperbolas parameterized for example by
\be{rindlert}
x = \chi\cosh\left(a\eta\right), \qquad t=\chi\sinh\left(a\eta\right),
\ee
where $a$ is the proper acceleration at the reference worldline  $\chi=1/a$ with proper time $\eta$.
(The notion of defining particles in a general curved spacetime is addressed in e.g. \cite{B&D,carroll}.
For the other special cases when the acceleration (i) is asymptotically uniform in the past/future see e.g. \cite{B&D,carroll,Mann:Villalba:09}, or (ii) asymptotically zero in the past but asymptotically uniform in the future and see e.g. \cite{Ostapchuk:Lin:Mann:Hu:11}).
The transformation suggests that a suitable choice of coordinates for uniformly accelerated observers are $(\eta,\chi)$ which are known as Rindler coordinates.
\begin{figure}[h]
\begin{center}
\includegraphics[width=.50\textwidth]{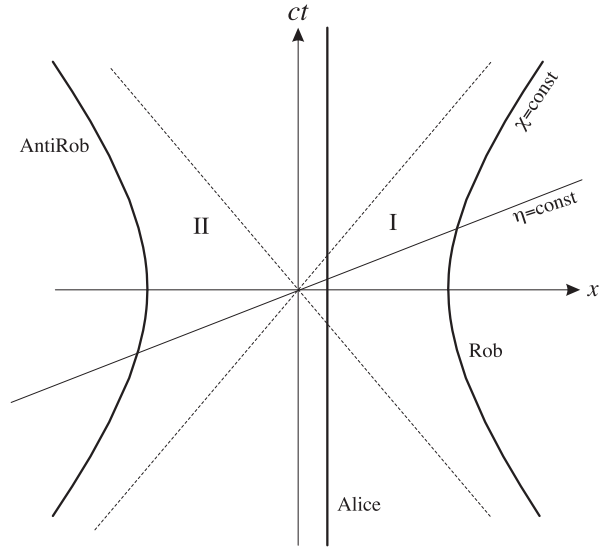}
\end{center}
\caption{ Rindler space-time diagram: lines of constant position $\chi$ are hyperbolae and all curves of constant  $\eta$ are straight lines that come from the origin. An uniformly accelerated observer Rob travels along a hyperbola constrained to either region $I$ or region $II$.}
\label{figbosons}
\end{figure}
The transformation in Eq.~(\ref{rindlert}) is defined in the region $|x|\geq t$ known as the (right) Rindler wedge I. When $\eta\rightarrow\infty$ then
$t/x=\tanh(a\eta)\rightarrow 1\Rightarrow x=t$.
Uniformly accelerated observers asymptotically approach the speed of light and are constrained to move in wedge I. Since the transformation does not cover all of Minkowski spacetime,
one must define a second region called (left) Rindler wedge II  by considering a coordinate transformation which differs from Eq. (\ref{rindlert}) by an overall sign in both coordinates.
Rindler regions I and II are causally disconnected, and the lines $x\pm t=0$ at 45 degrees
define the Rindler horizon, Fig.(\ref{figbosons}).

The metric in Rindler coordinates takes the form
$ds^2=(a^2\chi^2\,d\eta^2-d\chi^2)$
where $a^2\chi^2$ acts as an effective gravitational potential $g_{\eta\eta}(\chi)$ for the
Rindler observer's local redshift factor.
The Klein-Gordon equation in Rindler coordinates is
$(a^{-2}\partial_{\eta}^2 - \partial_{ln(a\chi)}^2)\phi=0$
and the solutions \cite{STakagi:86,brout,BSMA} are again plane waves, though now with logarithmic spatial
dependence $\ln\chi$, (compare with \Eq{eq:planem})
\numparts
\begin{eqnarray}\label{rindlerI}
u_{\om,I}&=& \frac{1}{\sqrt{4\pi\om}}\,e^{i(\eps (\om/a) \ln\chi - \om\eta)}
=\frac{1}{\sqrt{4\pi\Om}}\,\left( \frac{x-\eps t}{l_\Om}\right)^{i\eps\Om} \equiv u_{\Om,I}, \\
u^{*}_{\om,I}&=& \frac{1}{\sqrt{4\pi\om}}\,e^{-i(\eps (\om/a) \ln\chi - \om\eta)}
=\frac{1}{\sqrt{4\pi\Om}}\,\left( \frac{x-\eps t}{l_\Om}\right)^{-i\eps\Om} \equiv u^{*}_{\Om,I}.
\end{eqnarray}
\endnumparts
In the above $\om>0$, $\eps =1$ corresponds to modes propagating to the right along lines of constant $x-t$, and
$\eps=-1$ to modes propagating to the left along lines of constant $x+t$. In the second equality we
have introduced a positive constant $l_\Om$ of dimension length, and defined the dimensionless
positive constant $\Om = \om/a$. Some authors \cite{brout} choose to label the Rindler mode by
the (positive) frequency $\om$, while other authors \cite{STakagi:86,BSMA} label the modes by the (positive)
dimensionless quantity $\Om$. (Note that $-\infty < \eps\om/a = \eps\Om < \infty$ acts as the effective
wavevector for the Unruh modes, if one where to push the analogy with the inertial Minkowski modes \Eq{eq:planem}). Here we follow derivations from \cite{BSMA} and
throughout this work, it will be understood that a wavevector subscript $k$ on Minkowski modes
($u_{k,M}$, etc\ldots) takes values in the range $-\infty$ to $\infty$, while for Unruh modes
($u_{k,I}$, $u_{k,II}$ etc\ldots) it takes values from $0$ to $\infty$. 

The solutions $u_{k,I}$ and $u^{*}_{k,I}$ are identified as positive and negative frequency solutions, respectively,
with respect to the timelike Killing vector field $\partial_{\eta}$. These solutions have support only in the right Rindler wedge and
therefore are labeled by the subscript $I$. Note that they do not constitute a complete set of solutions.
The transformation which defines  Rindler region $II$
also gives rise to the same spacetime. However, the future-directed timelike Killing vector field
which in this case is given by $\partial_{(-\eta)}=-\partial_{\eta}$,  and the solutions are
\numparts
\begin{eqnarray}\label{rindlerII}
 u_{\om,II}&=& \frac{1}{\sqrt{4\pi\om}}\,e^{i(-\eps (\om/a) \ln(-\chi) + \om\eta)}
=\frac{1}{\sqrt{4\pi\Om}}\,\left( \frac{\eps t-x}{l_\Om}\right)^{-i\eps\Om} \equiv u_{\Om,II}, \\
 u^{*}_{\om,II}&=& \frac{1}{\sqrt{4\pi\om}}\,e^{-i(-\eps (\om/a) \ln(-\chi) + \om\eta)}
=\frac{1}{\sqrt{4\pi\Om}}\,\left( \frac{\eps t - x}{l_\Om}\right)^{i\eps\Om} \equiv u^{*}_{\Om,II},
\end{eqnarray}
\endnumparts
with support in region $II$. The solutions of region $I$ together with the solutions in region II form a complete set of orthonormal solutions.
Therefore, we can quantize the field in this basis as well,
\[\hat{\phi}=\int( u_{\Om,{I}}a_{\Om,I}+u_{\Om,{II}}a_{\Om,{II}}+h.c.)\,d\Om.\]
Since region $I$ is causally disconnected from region $II$, the mode operators in the separated
wedges commute $[a_{\Om,I}, a^\dag_{\Om',II}]=0$, etc.
 The vacuum state in the Rindler basis is $\left|0\right>_R=\left|0\right>^I\otimes\left|0\right>^{II}$ where  $a_{k,{I}}\left|0\right>^I=0$ and $a_{k,{II}}\left|0\right>^{II}=0$.
 Making use of the inner product we find the Bogoliubov transformations,
%
%
\Bea
a_{k,M} &=& \int(u_{k,M},u_{\Om,{I}})a_{\Om,{I}} + (u_{k,M},u_{\Om,I}^{\ast})a_{\Om,I}^{\dag} \no
  &+& (u_{k,M},u_{\Om, II})a_{\Om,II} + (u_{\Om,M},u_{\Om,II}^{*})a_{\Om,II}^{\dag}d\Om,
\Eea
where, for example, $(u_{k,M},u_{\Om,I})= -i\,\int(u^{*}_{k,M}\,\partial_t u_{\Om,I} - (\partial_t u_{\Om,I})\,u_{k,M}^{*})dx$.
Upon computing the inner products \cite{STakagi:86,BSMA} the above formula can be written as
\bea{Mink2Rindler}
a_{\omega,M} &=& \int^\infty_0 d\Omega\,
[(\alpha_{\omega\Omega}^R)^* ( \cosh(r_\Omega) a_{\Omega,I} - \sinh(r_\Omega) a^\dag_{\Omega,II}) \no
&+& (\alpha_{\omega\Omega}^L)^* (-\sinh(r_\Omega) a^\dag_{\Omega,I} + \cosh(r_\Omega) a_{\Omega,II})],
\eea
where
\Bea
\alpha_{\om\Om}^R &=& \frac{1}{\sqrt{2\pi\om}}\,(\om l)^{i\eps\Om}, \qquad
\alpha_{\om\Om}^L =\frac{1}{\sqrt{2\pi\om}}\,(\om l)^{-i\eps\Om},
\Eea
are the Bogoliubov coefficients for the massless case,
and $l$ is an overall constant of dimension length, independent of $\Om$ and $\eps$.

The Minkowski creation and annihilation operators result in an infinite sum of Rindler operators. An alternative basis for the inertial observers known as the Unruh basis can significantly simplify the transformations between inertial and uniformly accelerated observers. The Unruh modes $a_{\Omega,R}$, $a_{\Omega,L}$ are appropriately chosen linear combinations of right-moving and left-moving Rindler modes respectively such that they are analytic across both regions $I$ and $II$. That is, $u_{\Om,I}$ and  $u^{*}_{\Om,II}$ are both
proportional to $(x-\eps t)^{i\eps\Om}$ when $(-1)^{i\eps\Om}=(e^{i\pi})^{i\eps\Om} = e^{-\eps\pi\Om}$ is factored out of
the latter region $II$ mode.
The Unruh modes are given by the direct Bogoliubov transformation with region $I$ and $II$ Rindler modes
for each value of $\Om$ as
\bea{Unruh2Rindler}
a_{\Omega,R} &=& \cosh(r_\Omega) a_{\Omega,I} - \sinh(r_\Omega) a^\dag_{\Omega,II}, \no
a^\dag_{\Omega,L} &=& -\sinh(r_\Omega) a_{\Omega,I} + \cosh(r_\Omega) a^\dag_{\Omega,II},
\eea
where $\tanh(r_\Om) = e^{-\pi\Om}$.
Here $a_{\Omega,R}$ annihilates a right ($R$) moving Unruh mode traveling along lines of constant $x-t$ in both wedges $I$ and $II$, while $a_{\Omega,L}$ annihilates a left ($L$) moving Unruh mode traveling along lines of constant $x+t$, with
$[a_{\Omega,R}, a^\dag_{\Omega,R}]=1$, $[a_{\Omega,L},a^\dag_{\Omega,L} ]=1$ and all cross commutators vanishing.
In terms of mode functions, the Bogoliubov transformation from the Rindler to the Unruh modes is given by
\bea{eq:unruhmodes}
u_{\Om,R}     &=& \cosh({r})  u_{\Om,I} + \sinh({r})  u^{\ast}_{\Om,II}, \no
u^{*}_{\Om,L} &=& \sinh({r})  u_{\Om,I} + \cosh({r})  u^{\ast}_{\Om,II},
\eea
which are analytic in $x-t$ across both Rindler wedges $I$ and $II$.
Note that the sign of the momentum $k$ in region $II$ is opposite of that in region $I$,
but coupled with utilizing the complex conjugate of the region $II$ Rindler mode,
renders the resulting Unruh modes $u_{\Om,R}$ and $u^{*}_{\Om,L}$ right-movers (see \cite{carroll} Chap. 9.5 for further details).

The most general Unruh annihilation operator of purely positive Minkowski frequency is a linear combination of the two $R,L$ Unruh creation operators,
\be{oneUnruh}
a_{\Omega,U} = q_L a_{\Omega,L} + q_R a_{\Omega,R},
\ee
where $q_L$ and $q_R$ are complex numbers with $|q_L|^2+|q_R|^2=1$.
The introduction of the Unruh modes allows us to write the Minkowski annihilation operator \Eq{Mink2Rindler} as
a linear combination of only Unruh annihilation operators
\be{Mink2Unruh}
a_{\omega,M} = \int^\infty_0 d\Omega\, [(\alpha_{\omega\Omega}^R)^* a_{\Omega,R}
+ (\alpha_{\omega\Omega}^L)^* a_{\Omega,L}].
\ee
Hence, both Minkowski and Unruh annihilation operators annihilate the Minkowski vacuum,
i.e. $a_{\omega,M}|0\rangle_{\mathcal{M}}=0$, $a_{\Omega,R}|0\rangle_{\mathcal{M}}=0$, and $a_{\Omega,L}|0\rangle_{\mathcal{M}}=0$,
and therefore, the Unruh vacuum and the Minkowski vacuum coincide.


Using the direct Bogoliubov transformation between the Unruh and Rindler annihilation operators it is straight forward to show that (see e.g. \cite{WGUnruh:76,carroll,STakagi:86})
\begin{equation} \label{vacuumM}
\left|0_k\right>^{\mathcal{M}}=\frac{1}{\cosh(r)}\sum_n\tanh^{n}(r)\left|n_k\right>^{I}\left|n_k\right>^{II},
\end{equation}
where $\tanh r \equiv e^{-\pi\omega/a}$.
The vacuum state in the Rindler basis corresponds to a two mode squeezed state.   Since the accelerated observer is constrained to move in region $I$
one must trace over the states in (the causally disconnected) region $II$.  The density matrix of the Minkowski vacuum is given by  $\rho_0=\left|0_k\right>\left<0_k\right|^{\mathcal{M}}$
and therefore, the state in region $I$ corresponds to the following reduced density matrix
\begin{eqnarray*}
\rho_{I}&=&\frac{1}{\cosh^2(r)}\sum_{n}\tanh^{2n}(r)\left|n\right>^{I}\left<n\right|^{I},\\
&=&(1-e^{-2\pi\omega/a})\sum_{n}(e^{-2\pi\omega/a})^{n}\left|n\right>^{I}\left<n\right|^{I},
\end{eqnarray*}
which corresponds to a thermal state with temperature $T_{U}=\frac{a}{2\pi k_B}$ (where $k_B$ is the Boltzmann constant) proportional to the observer's acceleration.  The temperature is known as the Unruh temperature. This is the well known Unruh effect \cite{WGUnruh:76}:  the vacuum state as seen by inertial observers is a thermal state for observers in uniform acceleration.

\subsubsection{Unruh-DeWitt detectors}\label{Unruh-DeWitt_detectors}
In order to give a more physical interpretation to the Unruh effect, Unruh-DeWitt detectors were introduced \cite{WGUnruh:76,STakagi:86,brout,LCBCrispino:AHiguchi:GEAMatsas:08}. The detectors consist of a point-like system endowed with an internal structure which can be either a two level system or a harmonic oscillator \cite{PhysRevD.74.08503,PhysRevD.76.064008}. The detector moves in spacetime following a classical trajectory given by $x(\tau)$ where $\tau$ is the detector's proper time.  The detector couples  to the field locally via it's monopole moment.  Therefore, the interaction Hamiltonian for a harmonic oscillator detector is given by
\begin{equation}
\label{generalhamiltonian}
\hat{H}_{I}(\tau) = \int dk \lambda(\tau)(d\,e^{i\Omega\tau}+d^\dagger\,e^{-i\Omega\tau})\left(\hat{a}_k e^{i (k x(\tau)-\omega t(\tau))}+\hat{a}_k^\dagger e^{-i ( kx(\tau)-t(\tau))}\right),
\end{equation}
where $d$ and $d^{\dagger}$ are annihilation and creation operators for the internal degrees of freedom of the detector, $\Omega$ the frequency of the detector and $\omega=|k|$ the frequency of the field modes. The coupling function $\lambda(\tau)$  between the field and the detector can be chosen such that the interaction is switched on and off adiabatically. This removes transient effects.

Standard calculations \cite{B&D} consider a detector that follows either an inertial trajectory (say $x(t)= v t$ for a  detector with zero acceleration) or a uniformly accelerated one ($x=(t^2 + a^{-2})^{1/2}$).  At time $t=-\tau_0$ the field is in the vacuum state $|0\rangle_f$ and the detector in it's ground state $|0\rangle_d$. The detector is then turned on and the transition probability of the detector to an exited state at time $\tau_0$ is calculated using perturbation theory.  The transition rate per unit of proper time of the detector to a state with $n=1$ excitations at first order is given by
\begin{eqnarray}
\label{probtransrate}
\mathcal{P}=\sum_{n,\,\psi}\frac{1}{2\tau_0}\int_{-\infty}^{\infty} d\tau|A_n|^2,\quad
A_n= {}_d\!\langle 1_n| {}_f\!\langle \psi |H_{I}(\tau)|0\rangle_d|0\rangle_f,    
 \end{eqnarray}
where $|\psi\rangle_f$ is the final state of the field which is not observed, and hence, averaged over all possible outcomes.
This leads to the detector response function \cite{B&D,STakagi:86}
\[ F(\omega) = \lim_{s\downarrow 0}\lim_{\tau_0 \uparrow\infty}\frac{1}{2\tau_0}\int_{-\tau_0}^{\tau_0}\int_{-\tau'_0}^{\tau'_0} \,
         e^{-i\omega(\tau-\tau') -s|\tau|-s|\tau'}\,g(\tau,\tau'),
\]
taking $\lambda(\tau) = \exp(-s|\tau|)$, and defining $g(\tau,\tau')=G(x(\tau),x(\tau'))$ where
$G(x,x')= {}_f\!\langle 0|\,\hat{\phi}(x)\,\hat{\phi}(x')\,|0\rangle_f$ is the positive frequency Wightman function (WF),
and the field $\hat{\phi}(x)$ is given by the integral over all momentum of the
rightmost term in large parentheses in (\ref{generalhamiltonian}). In the above, the response function $F(\omega)$ is a consequence of Fermi's golden rule and depends only on the field and not on the structure of the detector, since the latter has been completely factored out \cite{B&D}. In general the WF is inversely proportional to the squared geodesic distance
$\sigma(\Delta x)=g_{\mu\nu} (x-x')^\mu (x-x')^\nu$ and hence is singular as $x\rightarrow x'$. This singular behavior is typically regularized by the ``$i\epsilon$" prescription which treats $F(\omega)$ as a contour integral in the complex $\Delta\tau=\tau-\tau'$ plane. For an inertial trajectory $x=vt$ one has \cite{B&D} $g(\Delta\tau) = -[2\pi(\Delta\tau -i\epsilon)]^{-2}$.
For $\omega>0$ the contour is closed in the lower half-plane for the integral to be convergent. However, since the pole in
$g(\Delta\tau)$ is in the upper half-plane, no contributions are obtained resulting in $F(\omega)=0$ as expected.
For the uniformly accelerated trajectory $x(\tau)=(t^2(\tau) + a^{-2})^{1/2}$ with $t(\tau) = a^{-1}\sinh(a\tau)$ one obtains
$g(\Delta\tau) = -[(4\pi/a) \sinh(a\Delta\tau/2-i\epsilon a)]^{-2}$. A contour integration \cite{B&D} now picks up contributions from the poles along the negative imaginary axis, leading to a response function proportional to
$\omega/(\exp(2\pi\omega/a)-1)^{-1}$. The appearance of the Planck factor $(\exp(2\pi\omega/a)-1)^{-1}$ indicates that the equilibrium reached between the accelerated detector and the field $\hat{\phi}$ in the vacuum state $|0\rangle_f$ is identical to the case when the detector remained unaccelerated, but was immersed in a bath at (Unruh) temperature $T_U = a/(2\pi k_B)$. Transition rates for Unruh-DeWitt detectors in curved spacetime have been considered in \cite{Louko2008}.
We will see in the next section that such accelerated detectors can be used to extract entanglement from the field.

Note that since the detector is point-like it couples with the same strength
to every frequency field mode. However this is not a very realistic situation. A more physical situation corresponds to a detector which has a spatial profile \cite{STakagi:86,Schlicht20044647,Louko2006}. Such detectors couple to a distribution of field modes which depends on the specifics of the profile. The use of detectors with infinitesimal, though non-zero, spatial extent can be utilized  to give a more physical interpretation of the mathematically formal ``$i\epsilon$" prescription used above in the computation of the detector response function. In brief, the use of infinitesimal sized detectors smears out the field $\hat{\phi}(x)$ along the detector trajectory, thus regularizing the singular behavior of the WF as $x\rightarrow x'$. Following Takagi (see \S 3.2 of \cite{STakagi:86}), one can consider the instantaneous non-rotating rest frame of the detector (the Fermi-Walker (FW) frame, see \S13.6 of \cite{MTW}). The coordinates transformation
$t(\tau,\zeta) = (a^{-1} + \zeta)\sinh(a\tau)$ and $x(\tau,\zeta) = (a^{-1} + \zeta)\cosh(a\tau)$ with metric
$ds^2 = (1+a\zeta)^2 d\tau^2 - d\zeta^2$, describes the observer's ``local laboratory" where the detector sits at the origin $\zeta=0$ of the local spatial coordinates. For a ``rigid" detector, its monopole field ${\cal{M}}(x(\tau,\zeta))$ (the term in leftmost parentheses in (\ref{generalhamiltonian}) can be written in a factorized form
${\cal{M}}(x(\tau,\zeta)) = \lambda(\tau)\,M(\tau)\,f(\zeta)$.
The detector-field interaction is now given by $H_{int}=\int d\tau \int d\zeta {\cal{M}}(x(\tau,\zeta))\,\hat{\phi}(x(\tau,\zeta))$
which is now of the form $\lambda(\tau)\,M(\tau)\,\hat{\phi}'(\tau)$, where
$\hat{\phi}'(\tau)=\int d\zeta f(\zeta)\,\hat{\phi}(x(\tau,\zeta)$ is the smeared out quantum field.
Using the mode expansion $\hat{\phi}(x(\tau)) = \int dk \, a_k U_k(x(\tau))$ with plane waves
$U_k(x(\tau))=e^{i k_\mu x^\mu(\tau)}$ allows the smeared field to be written as
$\hat{\phi}'(\tau)=\int dk\,a_k U_k(x(\tau)) \tilde{f}_k(\tau) + h.c.$ Here
$\tilde{f}_k(\tau) = \int d\zeta e^{i k_\mu [x^\mu(\tau,\zeta)-x^\mu(\tau)]}$
$= \int d\zeta e^{i k_\mu [\partial x^\mu(\tau,\zeta)/\partial \zeta^j]\,\zeta^j}$ are wavepackets that can be effectively
replaced with $\tilde{f}_k\approx e^{-\epsilon\omega_k/2}$ (upon substitution of the coordinate transformations to the FW frame)
where $\epsilon$ is a small positive quantity on
the order $1/a$. This latter form of $\tilde{f}_k$ is just the ``$i\epsilon$" prescription, and shows the advantage of
utilizing smeared fields/wavepackets to regularize divergent quantities, as well as lending physical interpretation to mathematical procedures. Work in progress shows that a uniformly accelerated detector with a Gaussian spatial profile naturally couples to a peak distribution of Rindler modes \cite{Ant}. Such a detector also naturally couples to distributions of Unruh modes which in this case yields a frequency distribution with two peaks corresponding to left and right moving Unruh modes.
This detector model will help in the understanding, from  physical perspective, the nature of Unruh modes which have been used to analyze the degradation of entanglement in non-inertial frames.

\subsubsection{Moving cavities}\label{Moving_cavities}
As mentioned previously, in order to preform quantum information tasks observers need to have access to the state and therefore, localizing field modes in space and time becomes relevant. This can be achieved by confining quantum fields in cavities which can move in spacetime \cite{TGDownes:IFUentes:TCRalph:10,DEBruschi:JLouko:IFuentes:12,NFriis:DEBruschi:JLouko:IFuentes:12,NFriis:ARLee:DEBruschi:JLouko:10,DEBruschi:Dragan:Lee:JLouko:IFuentes:12,NFriis:IFuentes,Friss:Hubber:Bruschi:IFuentes:12}. To confine a massless field within a cavity  of length $L$  appropriate Dirichlet boundary conditions must be imposed on the solutions of the Klein-Gordon equation given by (\ref{eq:planem}) at the cavity mirrors placed at  $x_{l}$ and $x_{r}=x_{l}+L$.  The normalized field solutions in this case correspond to sine functions
\begin{eqnarray}\label{icmodes}
U_{n,M}&=& \frac{1}{\sqrt{n\pi}}\sin\left(\omega_n[x-x_l]\right)e^{-i\omega_n t},
\end{eqnarray}
where $n\in N$ labels the energy states which now have a discrete spectrum given by $\omega_n=n\pi/L$. The Killing vector field is $\partial_t$ which classifies $U_{n,M}$ and $U^{\ast}_{n,M}$ as positive and negative frequency solutions. The normalization of the solutions gives $[U_{n,M},U_{m,M}]=[U^{\ast}_{n,M},U^{\ast}_{m,M}]=\delta_{nm}$ and the mix products vanish. The field is therefore,
\[\hat{\phi}=\sum_n(U_{n,M}a_{n,M}+U_{n,M}^*a_{n,M}^{\dag}),\]
where creation and annihilation operators satisfy standard bosonic commutation relations. We have used capitalized letters for the cavity modes to distinguish them from global modes.

We can also consider the field confined to a cavity in uniform acceleration by imposing uniformly accelerating boundary conditions to the modes in Eq.~(\ref{rindlerI})  at $\chi=\chi_r$ and $\chi=\chi_l$. Here we considered Rindler coordinates $(\eta,\chi)$ which are a suitable choice for this case once more. We consider the cavity to move in region $I$ without loss of generality.  The cavity is constructed such that it has a constant proper length as measured by a co-moving observer.  Therefore, the mirrors  move with different proper acceleration $A_r=1/x_r$ and $A_l=1/x_l$.
Taking advantage of the invariance under the boost Killing vector $\partial_\eta$ during acceleration we write 
  \begin{eqnarray}\label{rcmodes}
U_{n,R}&=& \frac{1}{\sqrt{n\pi}}\sin\left(\Omega_n[\chi-\chi_l]\right)e^{-i\omega_n \eta},
\nonumber
\end{eqnarray}
where $\Omega_n=n\pi/L'$ with $L'=\chi_r-\chi_l = a^{-1}\ln(1 + a L)$.
For the case $a L/c^2 \ll 1$ we have $L'\approx L$, and the small difference of the acceleration at the ends of the mirror can be ignored with respect to the acceleration at the center of the cavity.
Since we are only interested in the field within the mirrors, we dropped the index $I$ and instead included $R$ to denote the modes of the cavity, which is undergoing uniform acceleration.
The quantum field inside the cavity as seen by a co-moving observer is given by $\hat{\phi}_R(\eta,\xi)=\sum_n(U_{n,R}(\eta,\xi) a_{n,R}+U_{n,R}^*(\eta,\xi) a_{n,R}^{\dagger})$ where $a_{n,R}^{\dagger}$ and $a_{n,R}$ are once more creation and annihilation operators with $[a_{n,R}, a^{\dagger}_{n^{\prime},R}]=\delta_{n,n^{\prime}}$. The ground state, in this case, is defined by $a_{n,R}|0\rangle^R=0$, $\forall n$. Note that the Rindler coordinates completely cover the region inside the cavity and that the horizon always lies outside the cavity for all values of $a$. We assume the cavity's mirrors to be perfectly reflecting therefore, the states inside the cavity it will remain the same for all times \cite{TGDownes:IFUentes:TCRalph:10}. This is in agreement with Schutzhold and Unruh's \cite{Schutzhold:Unruh:2005} comment  concerning how the non-transmissive cavity protects the state from Unruh radiation. If one prepares a state, for example a pure state, as long as the cavity remains either inertial or in uniform acceleration, the state remains unchanged, i.e. pure.

\begin{figure}[h]\label{fig:alice}
\begin{center}
\includegraphics[width=.50\textwidth]{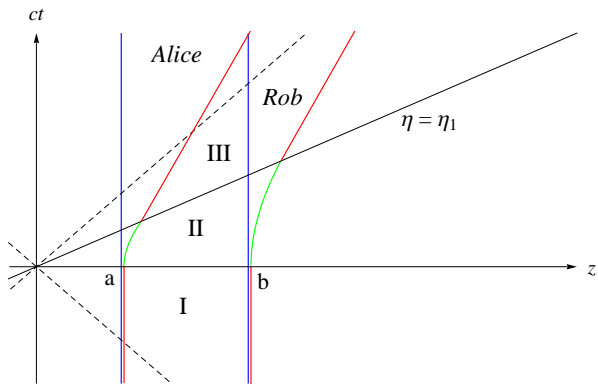}
\end{center}
\caption{Non-uniformly accelerated trajectory from initial constant velocity to uniformly accelerated motion.}
\label{non-uniform_traj}
\end{figure}

More interesting is to consider the case in which the cavity is placed in a spaceship initially at rest which uniformly accelerates for a finite amount of time and finally moves at constant velocity, as illustrated in figure \ref{fig:alice}.
This trajectory is interesting since it is more realistic (the uniformly accelerated trajectory assumes constant acceleration from past infinity to future infinity). Further, general travel scenarios can be crafted by considering sequences of it, and therefore it is called the basic building block trajectory\cite{DEBruschi:JLouko:IFuentes:12}. The modes inside an cavity initially at rest will be affected by the non-uniform accelerated motion giving rise to particle creation and entanglement  \cite{DEBruschi:JLouko:IFuentes:12,NFriis:DEBruschi:JLouko:IFuentes:12,NFriis:ARLee:DEBruschi:JLouko:10,DEBruschi:Dragan:Lee:JLouko:IFuentes:12,NFriis:IFuentes,Friss:Hubber:Bruschi:IFuentes:12},. It is possible to find Bogoliubov transformations between the modes of the inertial cavity and the modes of the cavity after any given travel scenario.  The new modes will be given by
\begin{equation}
a_{m,F}=\sum_n{\alpha^{\ast}_{mn}a_{n,M}-\beta^{\ast}_{mn}a^{\dagger}_{n,M}},
\end{equation}
where $F$ denotes the modes in the final region, and $\alpha_{mn}=(U_{m,F},U_{n,M})$ and $\beta_{mn}=-(U_{m,F},U^{\ast}_{n,M})$ are Bogoliubov coefficients. To study the above case in more detail it is convenient to work in the covariant matrix formalism which will be introduced in the section on quantum information. In the simple case of the building block trajectory the coefficients are given by
\numparts
\begin{eqnarray}\label{explicit:alpha:and:beta}
\alpha^{B}_{mn}=\frac{1}{L}\sqrt{\frac{n}{m}}F_{mn}+\frac{1}{\ln(\frac{x_r}{x_l})}\sqrt{\frac{m}{n}}G_{mn},\\
\beta^{B}_{mn}=\frac{1}{L}\sqrt{\frac{n}{m}}F_{mn}-\frac{1}{\ln(\frac{x_r}{x_l})}\sqrt{\frac{m}{n}}G_{mn},
\end{eqnarray}
\endnumparts
where
\begin{eqnarray*}
F_{mn}: &=&\int_{L_0}^{R_0}dx\sin\left(\omega_n(x-L_0)\right)\sin\left(\Omega_m\ln(\frac{x}{L_0})\right)\nonumber\\
G_{mn}: &=&\int_{L_0}^{R_0}\frac{dx}{x}\sin\left(\omega_n(x-L_0)\right)\sin\left(\Omega_m\ln(\frac{x}{L_0})\right).
\end{eqnarray*}
Here $B$ stands for basic building block.  These coefficients are difficult to handle both numerically and analytically. However, in the case where $h=aL$ is small, the terms in the integrand can be expanded in a Maclaurin series such that, to second order, the coefficients take a simple form  \cite{DEBruschi:JLouko:IFuentes:12,NFriis:DEBruschi:JLouko:IFuentes:12,NFriis:ARLee:DEBruschi:JLouko:10}.

\subsubsection{Curved spacetime}\label{Curved_spacetime}
For a general curved space metric $g_{\mu\nu}(x)$ the d'Alambertian (see section \ref{Global_fields}) gives rises to a more complicated, and very likely non-separable, Klein-Gordon equation. However, in some special cases solutions can be found \cite{B&D}.  An example is a $(1+1)$-dim expanding Robertson-Walker universe. This spacetime does not admit a global Killing vector field and therefore, it is not possible to define particles globally. This is also true for most spacetimes.  However, the Robertson-Walker universe has a special property in that it is asymptotically flat in the past and future infinity regions where timelike Killing vector fields can be defined and employed to distinguish positive and negative solutions to the Klein-Gordon equation. We will consider that a scalar field living in this spacetime is in the vacuum state as seen by observers in the past infinity region, and show that the state will be populated with particles in the future infinity.

The spacetime of a Robertson-Walker universe in $(1+1)$-dim is given by
$ds^2=dt^2-a^2(t)d\chi^2$
where the spatial sections of the spacetime are expanding (or contracting) uniformly according to the function $a^2(t)$. Considering the infinitesimal coordinate transformation
$d\eta=dt/a(t)$
and defining $a^2(t)=c^2(\eta)$ we obtain the metric
$ds^2=c^2(\eta)(d\eta^2-d\chi^2)$.
Consider $c(\eta)=1+\epsilon(1+\tanh(\sigma\eta))$ where $\epsilon$ and $\sigma$ are constants. This describes a toy model of a universe undergoing a period of smooth expansion. The parameter $\epsilon$ is known as the expansion volume and $\sigma$ is the expansion rate.
In the limit $\eta\rightarrow-\infty$ the metric is $ds^2=(d\eta^2-d\chi^2)$ and in the limit $\eta\rightarrow\infty$ then $ds^2=(1+2\epsilon)(d\eta^2-d\chi^2)$. The metric is flat in these regions where the vector field $\partial_{\eta}$ has Killing properties.

We now consider the massive Klein-Gordon equation $(\square+m^2)\phi=0$  in the Robertson-Walker spacetime described above where $m$ is the mass of the field.
The metric tensor $g_{\mu\nu}$ has components
\[g_{\mu\nu}=c(\eta)\left(\begin{array}{cc} 1&0\\0&-1\end{array}\right),\]
therefore, $g=det(g_{\mu\nu})=-c^2(\eta)$ and
the Klein-Gordon equation takes the form
\be{CS0}
((\partial_{\eta}^2-\partial_{\chi}^2)+c(\eta)m^2)\phi=0.
\ee
Exploiting spatial translational invariance we separate the solutions into
\[u_k=\frac{1}{\sqrt{2\pi\omega}}e^{ik\chi}\xi_k(\eta),\]
so that the equation becomes
\be{CS1}
\partial^2_{\eta}\xi_{k}(\eta)+(k^2+c(\eta)m^2)\xi_{k}(\eta)=0.
\ee
This equation can be solved in terms of two hypergeometric functions. We find two types of solutions \cite{B&D}
\numparts
\begin{eqnarray}
\label{CS2}
u^{(1)}_{k}(\eta,\chi)&=&\frac{1}{\sqrt{4\pi\omega_{in}}}e^{ik\chi-i\omega_{+}\eta-\left(\frac{i\omega_{-}}{\sigma}\right)\ln2\cosh(\sigma\eta)} {_2}F_{1}(\alpha,\beta,\gamma_{1},\delta_{+}),\\
u^{(2)}_{k}(\eta,\chi)&=&\frac{1}{\sqrt{4\pi\omega_{out}}}e^{ik\chi-i\omega_{+}\eta-\left(\frac{i\omega_{-}}{\sigma}\right)\ln2\cosh(\sigma\eta)} {_2}F_{1}(\alpha,\beta,\gamma_{2},\delta_{-}),
\end{eqnarray}
\endnumparts
where the constants above are defined as
\be{CS3}
\fl
\alpha = 1+\frac{i\omega_{-}}{\sigma},\quad
\beta=\frac{i\omega_{-}}{\sigma},\quad
\gamma_{1}=1-\frac{i\omega_{in}}{\sigma},\quad
\gamma_{2}=1+\frac{i\omega_{out}}{\sigma}\quad
\delta_{\pm}=\frac{1}{2}(1\pm\tanh(\sigma\eta)),
\ee
and the frequencies
$\omega_{in}=[k^2+m^2]^{\frac{1}{2}}$,
$\omega_{out}=[k^2+m^2(1+2\epsilon)]^{\frac{1}{2}}$ and
$\omega_{\pm}=\frac{1}{2}(\omega_{out}\pm\omega_{in})$.
We note that in the limit $\eta\rightarrow-\infty$ the first solution becomes
\[u_{k}^{(1)}\rightarrow\frac{1}{\sqrt{4\pi\omega_{in}}}e^{ik\chi-i\omega_{in}\eta},\]
while in the case $\eta\rightarrow+\infty$ the second solution is
\[u_{k}^{(2)}\rightarrow\frac{1}{\sqrt{4\pi\omega_{out}}}e^{ik\chi-i\omega_{out}\eta}.\]
One can see that in these limits the asymptotic solutions $u_{k}^{(1)}$ and $u_{k}^{(2)}$ to the Klein-Gordon equation are plane waves which can then be associated with positive mode solutions. The negative mode solutions correspond to $u_{k}^{1 *}$ and $u_{k}^{2 *}$. Since $u^1_{k}$ is associated with a plane wave at $\eta\rightarrow-\infty$ (past infinity) we call these solutions in-waves  $u_{k}^{(1)}\equiv u_{k}^{(in)}$.  The solutions $u_{k}^{(2)}\equiv u_{k}^{(out)}$ which are associated with plane waves at $\eta\rightarrow+\infty$ (future infinity) will be called out-waves.

Using the linear transformation properties of hypergeometric functions one can write $u_{k}^{(in)}$ in terms of $u_{k}^{(out)}$. This is easier than calculating the Bogoliubov coefficients via brute force direct integration using the inner product.
One obtains
\be{CS4}
u_{k}^{(in)}(\eta,\chi)=\alpha_{k}u_{k}^{(out)}(\eta,\chi)+\beta_{k}u_{k}^{(out *)}(\eta,\chi),
\ee
where
\numparts
\begin{eqnarray}\label{CS5}
\alpha_{k}&=&\left(\frac{\omega_{out}}{\omega_{in}}\right)^{\frac{1}{2}}\frac{\Gamma\left(1-\frac{i\omega_{in}}{\sigma}\right)\Gamma\left(-\frac{i\omega_{out}}{\sigma}\right)}{\Gamma\left(-\frac{i\omega_{+}}{\sigma}\right)\Gamma\left(1-\frac{i\omega_{+}}{\sigma}\right)},\\
\beta_{k}&=&\left(\frac{\omega_{out}}{\omega_{in}}\right)^{\frac{1}{2}}\frac{\Gamma\left(1-\frac{i\omega_{in}}{\sigma}\right)\Gamma\left(\frac{i\omega_{out}}{\sigma}\right)}{\Gamma\left(\frac{i\omega_{-}}{\sigma}\right)\Gamma\left(1+\frac{i\omega_{-}}{\sigma}\right)}.
\end{eqnarray}
\endnumparts
Here $\Gamma$ is the Gamma functions
with the properties $\Gamma(1+z) = z\,\Gamma(z)$,
$\Gamma(1-ix) = \pi/[\Gamma(ix)\sinh(\pi x)]$ and $|\Gamma(ix)|^2= \pi/[x\sinh(\pi x)]$ for $z$ complex and $x$ real.
From the above expressions we can read off the Bogoliubov coefficients  $\alpha_{kk'}=\alpha_k\delta_{kk'}$ and
$\beta_{kk'}=\beta_k\delta_{-kk'}$.
Therefore the transformation between annihilation operators yields
$a^{in}_{k}=\alpha_{k}^*a^{out}_{k}-\beta^*_{k}a_{-k}^{out \dag}.$
We now consider that the state of the field in the past infinity is the vacuum state,
\be{CS6}
{\ket{0}}^{in}=\bigotimes_{k=-\infty}^{\infty}{\ket{0}_{k}}^{in,}
\ee
and use the expression for the in-mode annihilation operator to calculate the state in the future infinity.
Since the transformation between in and out annihilation operators only mixes modes of frequency $k$ and $-k$, one finds that the state seen by observers in the future infinity is,
\be{CS7}\ket{0}_{k}^{in}=\sqrt{1-\gamma}\sum_{n}\gamma^n\ket{n}_{k}^{out}\ket{n}_{-k}^{out}.
\ee
where, \[\gamma=\frac{\sinh^2(\pi\omega_{-}/\sigma)}{\sinh^2(\pi\omega_{+}/\sigma)}.\]
The vacuum state from the perspective of observers in the remote past has particles in the remote future. Due to the expansion of the universe there has been particle creation.
The state $\ket{0}^{in}$ is (once again) a two-mode squeezed state, and
the particle creation process just described has a strong analogy with the quantum optical process of spontaneous parametric down conversion  if one describes only the signal photons emerging from the end of a nonlinear crystal, while ignoring the idler photons \cite{YurkePotasek87,AlsingMcMahonMilburn04}. (Note: here the laser pump acts as as the ``source" driving the expansion of the universe).
Before reviewing work on entanglement in all of the above scenarios, we will introduce the basic tools to quantify it.

\subsection{Quantum information}\label{Quantum_information}

The main aim of quantum information is to learn how to store, process and read information using quantum systems.  In this section we will briefly revise basic concepts of quantum entanglement   for pure and mixed states which are considered to be key resources in quantum information.
\subsubsection{Entanglement}\label{Entanglement}
 Entanglement is a quantum property which is a consequence of the superposition principle and the tensor product structure of the Hilbert space. The pure bi-partite case is well understood.
The state of two particles A and B is a vector in a $(d_a\times d_b)$-dimensional Hilbert space $\mathcal{H}_{ab}=\mathcal{H}_a\otimes \mathcal{H}_b$.  The space $\mathcal{H}_{ab}$ is the tensor product of the subspaces $\mathcal{H}_a$ and $\mathcal{H}_b$ of each particle.  An pure state element of the space $\mathcal{H}_{ab}$ is written as $\ket{\psi_{ab}}=\sum_{i,j} A_{ij}\ket{i}_a\otimes\ket{j}_b$.

A state $\ket{\psi_{ab}}\in \mathcal{H}_a\otimes \mathcal{H}_b$ is separable if $\ket{\psi_{ab}}=\ket{\phi}_a\otimes\ket{\varphi}_b$. Separable states can be prepared by local operations and classical communication. This means that observers manipulate each particle independently by making measurements, or applying unitary transformations of the form  $\ket{\psi_{ab}}=U_a\otimes U_b \,\ket{\phi}_a\otimes\ket{\varphi}_b$ where $U_a$ and $U_b$ are unitaries acting on particles A and B, respectively. The observers are also allowed to exchange classical information.  If the state is not separable then it is entangled. An entangled state cannot be prepared by local operations and classical communication, observers must make global operations on the systems. To determine whether or not a general pure state $\ket{\psi_{ab}}=\sum A_{ij}\ket{i}_a\ket{j}_b$ is entangled we consider the following theorem:
\begin{quote}
\textit{Schmidt decomposition}: Let $\mathcal{H}_a$ and $\mathcal{H}_a$ be Hilbert spaces of dimension $d_a$ and $d_b$, respectively.  For any vector $\ket{\psi_{ab}}\in \mathcal{H}_a\otimes \mathcal{H}_b$ there exists a sets of orthonormal vectors $\{\ket{j}_a\}_{a=1,\ldots,d_a}\!\subset\!\mathcal{H}_a \mbox{ and } \{\ket{l}_b\}_{b=1,\ldots,d_b}\!\subset\!\mathcal{H}_b$  such that we can write
$\ket{\psi_{ab}}=\sum_{i=1,\ldots,d}\lambda_i\ket{i}_a\ket{i}_b$.
The Schmidt coefficients $\lambda_i$ are non-negative scalars such that $\sum_{i} \lambda^2_i=1$, and
the sum runs over the minimum dimension $d=\textrm{min}(d_a,d_b)$ of the two Hilbert spaces.
\end{quote}

This special basis is called the Schmidt basis \cite{MikeandIke:00,Peres:95}.
Note that the correlations between systems A and B are now made explicit. For example,
if $\lambda_i=1$ and $\lambda_{j\ne i}=0$  then the state is separable. In the case that all the $\lambda_i$'s are equal and $\lambda_{i}=1/\sqrt{d}$ then the state is maximally entangled.
It is clear that the distribution of the Schmidt coefficients determine how entangled the state is. Therefore, to quantify entanglement in the pure bi-partite case we need a monotonous and continuous function of the $\lambda_i$'s such that
\begin{enumerate}
\item $S(\lambda_i)=0$ for separable states,
\item $S(\lambda_i)=\log(d)$ for maximally entangled states.
\end{enumerate}
Considering the density matrix $\rho_{ab}=\ket{\psi_{ab}}\bra{\psi_{ab}}$ and it's reduced density matrix $\rho_b=Tr_a(\rho_{ab})$, we find that the von Neumann entropy
\begin{eqnarray*}
S(\rho_b)&=&-Tr\left(\rho_b\log_2\rho_b\right)=-\sum_i|\lambda_i|^2\log_2|\lambda_i|^2,
\end{eqnarray*}
quantifies the entanglement between system A and B. We observe from the Schmidt decomposition that it is equivalent to trace over either system A or B and therefore, $S(\rho_a)=S(\rho_b)$. The von Neumann entropy of a pure state is $S(\rho_{ab})=0$.

Quantifying entanglement in the mixed case is more involved since now there is no analog to the Schmidt decomposition.
However, it is still possible to define a separable mixed state.
A bipartite mixed state is separable if we can write its density matrix as
$\rho_{ab}=\sum_i p_i\rho_a^i\otimes\rho_b^i$ where $\sum_i p_i=1$.

To determine whether or not a general mixed state is entangled it is convenient to define the partial transpose of a density matrix. Consider the general bipartite mixed state
\begin{equation}
\rho_{ab}=\sum_{ijkl}C_{ijkl}\,|i\rangle_a  |j\rangle_b \, {}_b\langle k|  {}_a\langle l|
\equiv\sum_{ijkl} C_{ijkl}\,|i\rangle_a{}_a\langle l| \otimes |j\rangle_b{}_b\langle k|. \nonumber
\end{equation}
The partial transpose $\rho^{PT}_{ab}$ of  $\rho_{ab}$ is (transposing on system $a$)
\begin{equation}
\rho_{ab}^{PT}=\sum_{ijkl}C_{ljki}\,|i\rangle_a|j\rangle_b{}_b\langle k|_a\langle l|, \nonumber
\end{equation}
or equivalently (transposing on system $b$)
\begin{equation}
\rho_{ab}^{PT}=\sum_{ijkl}C_{ikjl}|i\rangle_a|j\rangle_b{}_b\langle k|_a\langle l|. \nonumber
\end{equation}
Since the partial transpose of a separable state has positive eigenvalues it is possible
to construct a bipartite separability criterion:
If the eigenvalues of
$\rho_{ab}^{PT}\geqslant 0$ then $\rho_{ab}$ is separable. However, this criterion is only sufficient for $2\times2$ and $2\times3$ systems \cite{MHorodecki:PHorodecki:RHorodecki:98}. For systems of higher dimension the criterion is only necessary, meaning that there are (bound) entangled states with positive partial transpose. Such states are known as bound entangled states.

Adding up the negative eigenvalues gives an estimate of how entangled a state is.  Therefore, we will now define the negativity and logarithmic negativity \cite{GVidal:RFWerner:11,MBPlenio:SVirmani:07} which are two entanglement monotones.
The negativity of a density matrix $\rho_{ab}$ is defined as the sum of the absolute values of the negative eigenvalues of the partial transpose $\rho_{ab}^{PT}$,
\[N(\rho_{ab}):=\frac{\|{\rho^{PT}_{ab}}\|-1}{2},\]
where $\|\cdot\|$ denotes the trace norm $\|X\|:=Tr\left[\sqrt{X^{\dag}X}\right]$.
The logarithmic negativity of a density matrix $\rho_{ab}$ is defined as $E_N(\rho_{ab}):=\log_2\|\rho_{ab}^{PT}\|.$

\subsubsection{Covariant Matrix formalism}\label{Covariant_Matrix_formalism}
The covariant matrix formalism is usually employed in quantum optics \cite{JLaurat:GKeller:JAOliveiraHuguenin:CFabre:TCoudreau:ASerafini:GAdesso:FIlluminati:05}.  It is a framework that involves simple mathematical tools which are applicable to systems consisting of a finite number of harmonic oscillators.
While in general computing entanglement can be very involved, it has been shown that in the case of Gaussian states (such as coherent, squeezed and thermal states)  the covariant matrix formalism can be employed to produce computable measures of quantum information \cite{GAdesso:07}. Gaussian states are described by quasi-probability distributions of Gaussian shape in phase space \cite{AdessoIlluminati2005}. This technique is very useful in relativistic quantum information since Bogoliubov transformations map Gaussian states back onto Gaussian states \cite{NFriis:IFuentes,GAdesso:IFuentesSchuller:MEricsson:07,adessosamy}. As long as one considers initially Gaussian states and a finite number of field modes it is possible to calculate entanglement and other interesting quantities using the measures developed in this framework. In the standard Fock space description of quantum field theory quantum states correspond to density matrices which contain all the information pertaining to the state. In the covariant matrix formalism the notion of density matrix is replaced by a real symplectic covariant matrix $\sigma$ defined by
\begin{equation}
    \sigma_{ij}\,=\,\left\langle\,\mathrm{X}_{i}\mathrm{X}_{j}\,+\,\mathrm{X}_{j}\mathrm{X}_{i}\,\right\rangle\,
	   -\,2\,\left\langle\,\mathrm{X}_{i}\,\right\rangle\left\langle\,\mathrm{X}_{j}\,\right\rangle,
    \label{eq:covariance matrix}
\end{equation}
where the operators $\mathrm{X}_{i}$ are {the generalized} positions and momenta, i.e.
\numparts
\begin{eqnarray}
	\mathrm{X}_{(2n-1)}	&=	 \frac{1}{\sqrt{2}}(a_{n}\,+\,a_{n}^{\dagger})\,,
		\label{eq:position operators}\\[0.5mm]
	\mathrm{X}_{(2n)}	&=   \frac{-i}{\sqrt{2}}(a_{n}\,-\,a_{n}^{\dagger})\,,
		\label{eq:momentum operators}
\end{eqnarray}
\endnumparts
 and the index $n=1,2,3,\ldots$ labels the modes, and $\langle\,\mathcal{O}\,\rangle$ denotes the expectation value of the operator $\mathcal{O}$ with respect to the initial Gaussian state.
 The matrix $\sigma$ together with the vector of first moments
 $\langle\,\mathrm{X}_{i}\,\rangle$ completely characterizes all Gaussian states. However, the covariance matrix is sufficient to compute entanglement.
Unitary transformations in the Fock space are replaced in this formalism by symplectic transformations in phase space. A transformation $S$ is called symplectic, if it leaves the symplectic form $\Omega$ (a numerical matrix) invariant, i.e., $S\,\Omega\,S^{T}=\Omega$, where $[X_{i},X_{j}]=2i\Omega_{ij}$ and $(S_{ij})^{T}=(S_{ji})$.
The expression for the
symplectic representation of a Bogoliubov transformation in terms of its
general coefficients $\alpha_{mn}$ and $\beta_{mn}$ takes a simple form \cite{NFriis:IFuentes}. The matrix~$S$ is decomposed into $2\times2$ blocks ${\mathcal{M}}_{mn}$ as
\begin{equation}
    S=
    \left(\begin{array}{cccc}
        {\mathcal{M}}_{11}  &   {\mathcal{M}}_{12}  &   {\mathcal{M}}_{13}   &   \ldots  \\
        {\mathcal{M}}_{21}  &   {\mathcal{M}}_{22}  &   {\mathcal{M}}_{23}   &   \ldots  \\
        {\mathcal{M}}_{31}  &   {\mathcal{M}}_{32}  &   {\mathcal{M}}_{33}   &   \ldots  \\
        \vdots              &   \vdots              &   \vdots               &   \ddots
    \end{array}\right),
    \label{eq:Gaussian n-mode Bogo transformation}
\end{equation}
where the sub-blocks are given by
\begin{equation}
    \mathcal{M}_{mn}\,=\,
        \left(\begin{array}{cc}
            \Re(\alpha_{mn}\,-\,\beta_{mn})     &   \Im(\alpha_{mn}\,+\,\beta_{mn}) \\[1.5mm]
            -\Im(\alpha_{mn}\,-\,\beta_{mn})    &   \Re(\alpha_{mn}\,+\,\beta_{mn}) \\
        \end{array}\right)\,.
    \label{eq:M Bogo matrix}
\end{equation}
Here $\Re(z)$ and $\Im(z)$ denote the real part and imaginary part of the complex number $z$ respectively. The
transformed covariance matrix $\tilde{\sigma}$ is then simply obtained as $\tilde{\sigma}\,=\,S\,\sigma\,S^{T}\,.$
This transformation ensures that if $\sigma$ is Gaussian, $\tilde{\sigma}$ remains Gaussian.
Fortunately, for Gaussian states computable measures of entanglement have been developed \cite{GAdesso:07,AdessoSerafiniIlluminati2006}. For example,  the negativity between two modes is given by
 \begin{equation}
    \mathcal{N}=\max\{0,(1-\widehat{\nu}_{-})/2\widehat{\nu}_{-}\},
    \label{eq:negativity def}
\end{equation}
where $\widehat{\nu}_{-}$ is the smallest symplectic eigenvalue of the partial transpose matrix $\,\widehat{\sigma}_{kk^{\prime}}=T_{k^{\prime}}{\sigma}_{kk^{\prime}}\,T_{k^{\prime}}\,$,
where $T_{k^{\prime}}=\textrm{diag}\{1,1,1,-1\}$ partially transposes mode~$k^{\prime}$,
see Ref.~\cite{AdessoIlluminati2005}.

Computing the negativity between two modes $k$
and~$k^{\prime}$  involves first  tracing over all other modes.  Impressively, the partial trace over any subset of modes is computed  by
eliminating all rows and columns corresponding to all modes other than $k$
and~$k^{\prime}$. The smallest symplectic eigenvalue $\widehat{\nu}_{-}$ is then obtained by diagonalizing the matrix
$i\Omega\,\widehat{\sigma}_{kk^{\prime}}$  by a symplectic operation $D$, yielding the two eigenvalues
 $0\leq\widehat{\nu}_{-}\leq\widehat{\nu}_{+}$.
The state  ${\sigma}_{kk^{\prime}}$ is entangled if  $0\leq\widehat{\nu}_{-}<1$.

\section{Entanglement of global modes}\label{Entanglement_of_global_modes}

\subsection{Flat spacetime entanglement}\label{Flat_spacetime_entanglement}
In the Unruh effect the Minkowski vacuum is a thermal state for uniformly accelerated observers.
Since the Rindler regions $I$ and $II$ are causally disconnected, uniformly accelerated observers in one Rindler region (wedge)  have no access to information from the other Rindler region. Therefore, the state of the Rindler observer, adapted to his/her particular wedge by tracing out over the unaccessible wedge, is mixed.
The state appears more mixed  for observers with increasing
acceleration. By this we mean that the temperature associated to the state by a particular observer is higher for observers with larger proper accelerations. This has an effect on entanglement [3, 4]. Consider a pure entangled
Minkowski state of two field modes each of them labeled by a different Minkowski frequency.  We assume that two inertial observers, Alice and Bob,
are able to distinguish these modes and agree that the state is maximally entangled.
As typically considered in the literature, Alice and a Rindler observer in
region $I$, called Rob, also analyze the state. Since Rindler observers in Region $I$ must trace over the region
$II$ part of the state, the correlations of the state will be partially lost.
Since the vacuum itself exhibits greater noise for observers with higher accelerations one also expects
other states to appear more mixed and less correlated for higher accelerations. Therefore,
there is a degradation of entanglement due to the Unruh effect. Let us point out that
general region $I$ states are not necessarily thermal after tracing out over region $II$ states,
however, they are mixed. A thermal state is a particular type of mixed state in which the
fraction of the ensemble in each pure state is given by a Boltzmann distribution. In the
case of Gaussian states such as the vacuum, thermal states are obtained after tracing
over modes.
Calculations involving Minkowski states are very involved because a single Minkowski mode corresponds to an infinite  superposition (continuous) of
Rindler modes \eref{Mink2Rindler} making the trace operation highly non-trivial. It is therefore, more convenient from a mathematical point of view to analyze inertial states involving Unruh modes such as the Bell state,
\begin{equation}\label{maxent1}
\ket{\Psi}=\frac{1}{\sqrt2}\left(\ket{0_{k}}^{\mathcal{M}}\ket{0_{\Omega}}^{\mathcal{U}}+\ket{1_{k}}^{\mathcal{M}}\ket{1_{\Omega}}^\mathcal{U}\right),
\end{equation}
where ${\mathcal{M}}$ and $\mathcal{U}$ label Minkowski and Unruh states respectively with frequencies $k$ and $\Omega=\omega/a$.  This state was introduced in {\it Bruschi, et.al} \cite{BSMA}. For more details on the use of Unruh modes to analyze entanglement and the single mode approximation see the appendix, section \ref{appendix}.
For inertial observers the modes $k$ and $\Omega>0$ (which must be two distinguishable modes) are maximally correlated since the Bell state is maximally entangled. It is then interesting to investigate to what degree the state is entangled when described by observers in uniform acceleration.
In the simplest scenario shown in Fig. (\ref{figbosons}),  Alice is considered to be inertial and a uniformly accelerated observer Rob is introduced, who analyzes mode $\Omega$.  To study this situation, the states corresponding to Rob must be transformed into the appropriate basis, in this case, the Rindler basis.  We have already calculated the transformation for the vacuum state in \eref{Unruh2Rindler}, and with that in hand, we find the single particle Unruh state
\be{FSTE1}
a_{\Omega,U}^{\dag}|0_{\Omega}\rangle^{\mathcal{U}}
=\left|1_{\Omega}\right>^{\mathcal{U}}=\frac{1}{\cosh(r)}
\sum_n\tanh^{n}(r)\sqrt{n+1} \left|(n+1)_{\Omega}\right>_{I} \left|n_{\Omega}\right>_{II},
\ee
where we have chosen $|q_R|=1$ in Eq. (\ref{oneUnruh}).
Thus, the maximally entangled state seen by inertial Alice and accelerated Rob is
\bea{FSTE2}
\ket{\Psi}&=&\frac{1}{\sqrt{2}}\left|0_k\right>^{\mathcal{M}}\otimes\frac{1}{\cosh(r)}\sum_n\tanh^{n}(r)\left|n_{\Omega}\right>_{I}\left|n_{\Omega}\right>_{II} \no
&+&\frac{1}{\sqrt{2}}\left|1_k\right>^{\mathcal{M}}\otimes\frac{1}{\cosh(r)}\sum_n\tanh^{n}(r)\sqrt{n+1}\left|(n+1)_{\Omega}\right>_{I}\left|n_{\Omega}\right>_{II}.
\eea
Since Rob, with trajectory in region $I$, is causally disconnected from region $II$ we must take the trace over region $II$.
The density matrix for the Alice-Rob subsystem is
\be{FSTE3}
\rho_{AR}=\frac{1}{\cosh^2(r)}\sum_{n=0}^{\infty}\tanh^{2n}(r)\rho_n,
\ee
where
\begin{eqnarray*}
\fl \rho_n&=&\left|0_{k},n_{\Omega}\right>\left<0_{k},n_{\Omega}\right|+
\frac{\sqrt{n+1}}{\cosh(r)}\left(\left|0_{k},n_{\Omega}\right>\left<1_{k},(n+1)_{\Omega}\right|+\left|1_{k},(n+1)_{\Omega}\right>\left<0_{k},n_{\Omega}\right|\right)\\
\fl &+&\frac{n+1}{\cosh^2(r)}\left|1_{k},(n+1)_{\Omega}\right>\left<1_{k},(n+1)_{\Omega}\right|,
\end{eqnarray*}
involves only ${\mathcal{M}}$ and region $I$ states.
The entanglement between the modes can be quantified by the logarithmic negativity yielding $\mathcal{N}(\rho_{AR})=\log_2((1/2\cosh^2(r))+\Sigma)$ where
\be{FSTE4}
\Sigma=\sum_n\frac{\tanh^{2n}(r)}{2\cosh^2(r)}\sqrt{\left(\frac{n}{\sinh^2(r)}+\tanh^2(r)\right)^2+\frac{4}{\cosh^2(r)}}.
\ee

One observes that entanglement is degraded for observers in uniform acceleration and therefore, it is observer dependent. In the flat spacetime case, one can argue that this is an effect of Rob being non-inertial. Rob must be in a spaceship to be accelerated and energy must be supplied to accelerate the system. In the flat spacetime case, inertial observers play a special role and therefore,  a well defined notion of entanglement corresponds to the entanglement described from the inertial perspective. However, in curved spacetime different observers
describe a different particle content in the field which results in different degrees of entanglement in the field. In that case, there is no well-defined notion of entanglement.

Using Unruh rather than Minkowski states considerably simplifies the mathematical techniques involved in the analysis of entanglement. However, it introduces interpretational issues \cite{andrzej}.  The physical interpretation of Unruh states requires deeper understanding. Work in progress shows that finite size accelerated detectors naturally couple to peaked distributions of Unruh modes \cite{Ant}. Such detectors will provide further insights on the states considered in  Eq. (\ref{maxent1}). Currently, it is only possible to say that the work of  {\it Bruschi, et.al.} \cite{BSMA}  studies the entanglement of a family of maximally entangled states parameterized by the dimensionless $\Omega=\omega/a$ in non-inertial frames.
By fixing the physical frequency $\omega$ and changing $a$ one analyzes the entanglement in a family of states, all of which have the same frequency $\omega$ as seen by observers with different proper acceleration $a$.  An alternative view is that the entanglement corresponds to a family of states with different physical frequency $\omega$ as seen by the same observer moving with fixed proper acceleration $a$ \cite{PhysRevD.86.025026}.

The analysis of entanglement in non-inertial frames has been generalized in several directions and different quantum fields have been considered. For example, the work by {\it Adesso, et. al.} \cite{GAdesso:IFuentesSchuller:MEricsson:07} considers two-mode squeezed states in the inertial frame (instead of the Bell states analyzed above) allowing them to employ  the covariance matrix formalism which is only applicable for Gaussian states. The authors studied the case when both observers accelerate finding that entanglement vanishes at finite acceleration. This effect is also known in the literature as sudden death of entanglement \cite{Yu2009598}.  The effect is similar to that which occurs in parametric down conversion when a two-mode entangled state is feed into an amplification channel (see also \cite{Bradler2012}).  It was shown that the rate of entanglement degradation depends on the entanglement in the inertial frame. States with higher degrees of inertial entanglement degrade more in non-inertial frames. The degradation of entanglement has also been analyzed for inertial thermal states and non-maximally entangled states \cite{PhysRevA.77.024302}.

 Other types of correlations relevant to quantum information theory have also been studied in non-inertial frames. For example, it was found that classical correlations are conserved in case one observer is non-inertial \cite{Alice,PMAlsing:IFuentesSchuller:RBMann:TETessier:06} and degraded when both observers accelerate \cite{GAdesso:IFuentesSchuller:MEricsson:07}. More recently, the quantum discord \cite{PhysRevLett.88.017901,Henderson20016899}, a type of purely quantum correlations which are present even in separable states, was analyzed in non-inertial frames. {\it Datta} \cite{PhysRevA.80.052304} found that in a regime where there is no distillable entanglement there is a finite amount of quantum discord. Interestingly, in the limit of infinite acceleration the discord does not vanish.  Multi-partite entanglement has also been subject of study for observers in uniform acceleration \cite{PMAlsing:IFuentesSchuller:RBMann:TETessier:06,GAdesso:IFuentesSchuller:MEricsson:07}. The bi-partite entanglement between modes is degraded in non-inertial frames, however, three and four-partite correlations are created between modes in regions $I$ and $II$ \cite{GAdesso:IFuentesSchuller:MEricsson:07,GAdesso:IFuentesSchuller:09}. Multipartite entanglement has also been analyzed when the state contains tri-partite correlations in the inertial frame \cite{ASmith:RBMann:11}.

 The analysis of entanglement in non-inertial frames has been
 considered for different types of fields. For example, Dirac fields \cite{PMAlsing:IFuentesSchuller:RBMann:TETessier:06,PhysRevA.77.024302,ASmith:RBMann:11,PhysRevA.80.012314,PhysRevA.84.012337,PhysRevA.83.052306,PhysRevA.84.062111}, the electromagnetic field \cite{Ling20079025} and charged bosonic fields have been studied \cite{PhysRevD.86.025026}. The results for fermionic fields show that entanglement is degraded in non-inertial frames however, the entanglement remains finite in the infinite acceleration limit \cite{PMAlsing:IFuentesSchuller:RBMann:TETessier:06,PhysRevA.77.024302,PhysRevA.83.052306,Montero2011a}. Interestingly, for fermions the entanglement in the infinite acceleration limit does not violate Bell's inequalities \cite{PhysRevA.84.062111}.  Multipartite entanglement also presents interesting differences. In the bosonic case genuine tripartite entanglement is generated amongst the modes of Alice, Rob and Anti-Rob (an observer with uniform acceleration in region $II$) \cite{GAdesso:IFuentesSchuller:MEricsson:07}. These correlations increase with acceleration.
 However, in the fermionic case tripartite entanglement is always zero\cite{PMAlsing:IFuentesSchuller:RBMann:TETessier:06}.
 Recent work \cite{ASmith:RBMann:11} considers an inertial tripartite state which contains genuine tripartite correlations, as measured by the Svetlichny inequality \cite{Svetlichny:87}. These correlations persist for any finite acceleration for both GGHZ and MS three-qubit states provided the respective control parameters are appropriately chosen, vanishing only in the infinite acceleration limit. This indicates that tripartite entanglement and its associated nonlocal correlations are more robust to relativistic effects than their bipartite counterparts. It is important to mention that an ambiguity on the definition of entanglement between fermionic fields has been pointed out in \cite{CabanPodlaskiRembielinskiSmolinskiWalczak2005,PhysRevA.83.062323}. However, these results are subject of a present debate \cite{PhysRevA.85.016301,PhysRevA.85.016302}. (Note that in \cite{PMAlsing:IFuentesSchuller:RBMann:TETessier:06}, signs arising from the transposition of anti-commuting fermionic operators could always be absorbed into an explicit phase that was introduced, and hence into a redefinition of the fermionic creation operators).

 The main differences in the entanglement degradation for scalar and fermionic fields arise due to particle statistics
 \cite{PhysRevD.86.025026,PhysRevA.83.052306,Martín-Martínez2010a,Martín-Martínez2010b}
 and not to dimensionality of the Hilbert space.  In the fermionic case, a redistribution of entanglement between particle and antiparticle states seams to prevent the degradation of entanglement in the infinite acceleration limit \cite{PhysRevA.83.052306}. Such redistribution does not occur in the bosonic case \cite{PhysRevD.86.025026}.  Initially it was conjectured that the Hilbert space dimensionality played a role \cite{PMAlsing:IFuentesSchuller:RBMann:TETessier:06,PhysRevA.77.024302}. In the bosonic case the Hilbert space is infinite dimensional while in the case of Grassmann scalars, the lowest dimensional fermions, the fermionic Hilbert space is of dimension two. One might suspect that entanglement is degraded stronger in the bosonic case since for higher accelerations the bosonic harmonic oscillators become populated with increasing temperature while Grassmann scalars are always in mixed states which involve only two states. Considering higher dimensional fermions \cite{PhysRevA.80.042318} and bosonic fields with truncated Hilbert spaces \cite{Martín-Martínez2010a} showed that it is statistics rather than dimensionality which produces the observed differences in the degradation of entanglement in non-inertial frames. In the case of entanglement degradation for electromagnetic fields it was found that photon helicity entangled states do not degrade with acceleration \cite{Ling20079025}.

\subsection{Entanglement in curved spacetime.}\label{Entanglement_in_curved_spacetime}

General curved spacetimes do not have global timelike Killing vector fields and therefore it is not possible to define subsystems.
Without the notion of subsystem
it is not possible to study entanglement.  In spite of this, special cases have been analyzed. For instance,
black hole spacetimes where the spacetime is approximately flat at the horizon, and Robertson-Walker universes which have two asymptotically flat spacetime regions.
\subsubsection{Entanglement across black hole horizons}\label{Entanglement_across_black_hole_horizons}
The Schwarzchild spacetime of an eternal black hole describes the geometry of a spherical non-rotating mass $m$ \cite{B&D}. Considering only the radial component, the metric is
\begin{eqnarray*}
ds^2=(1-(2m/R))dT^2-(1-(2m/R))^{-1}dR^2.
\end{eqnarray*} where the horizon of the black hole is at $R=2m$. Considering the following coordinate change $R-2m=x^2/8m$, such that \[1-(2m/R)=x^2/8mR=\frac{(x^2/8m)}{(x^2/8m+2m)}=\frac{(Ax)^2}{(1+(Ax)^2)}\approx (Ax)^2\]
when $x\approx 0$ with $A=1/4m$. This means that $dR^2=(Ax^2)dx^2$. Therefore,  very close to the horizon $R\approx 2m$, the Schwarzschild spacetime can be approximated by Rindler spacetime
\begin{eqnarray*}
ds^2=-(Ax)^2dT^2+ dx^2,
\end{eqnarray*}
where the acceleration parameter $a=A^{-1}=4m$.

This means that, very close to the horizon of the black hole, we can consider Alice as being an inertial observer and falling into the black hole, while Rob escapes the fall by being a stationary, accelerated observer. If Alice claims that the state of the field is the vacuum state, then Rob detects a thermal state of the form
\begin{equation}\label{eq:bh}
\left|0_k\right>^{\mathcal{\mathcal{M}}}=\frac{1}{\cosh(q)}\sum_n\tanh^{n}(q)\left|n_k\right>^{in}\left|n_k\right>^{out},
\end{equation}
where $\cosh(q)=(1-e^{2m\pi\omega})^{-1/2}$ is a function of the mass of the black hole $m$. Here $in$ and $out$ denote the modes inside and outside the black hole. $ \left|0_k\right>^{\mathcal{\mathcal{M}}}$ is the state detected by Alice. If we then consider the case where two modes of the field in Alice's frame are maximally entangled, Rob will detect less entanglement between the modes due to the Hawking effect
\cite{Alice,GAdesso:IFuentesSchuller:09,Ahn2007368,Ahn2007202}. Using the covariant matrix formalism an inertial two-mode squeezed state  was analyzed \cite{GAdesso:IFuentesSchuller:09}. The entanglement between the modes is degraded for observers outside the black hole, however genuine four-partite entanglement between the modes inside and outside the black hole are created.  For low black hole masses entanglement vanishes and classical correlations also degrade. The analysis of entanglement degradation  has also been considered taking into account the distance of the observer to the event horizon \cite{Martín-Martínez2010c} by writing  the metric in terms of the proper time $\tau_0$ of a stationary observer placed at $r_0$,
\begin{eqnarray*}
ds^2=-(\tilde{A}x)^2d\tau_0^2+ dx^2
\end{eqnarray*}
which yields a modified acceleration $\tilde{A}=A/\sqrt{1-2m/r_0}=(4m\sqrt{1-2m/r_0})^{-1}$ giving rise to the same state as in Eq.~(\ref{eq:bh}) with a modified squeezing parameter $\tilde{q}$ given by $\cosh(\tilde{q})=(1-e^{2m\pi\omega/\sqrt{1-2m/r_0}})^{-1/2}$. This study has been also carried out in the fermionic case \cite{Martín-Martínez2010b,PhysRevA.80.042318}, and in higher dimensional black holes \cite{Ge2008}.

 \subsubsection{Entanglement in an expanding universe}\label{Entanglement_in_an_expanding_universe}

Our second curved space example is the Robertson-Walker universe \cite{B&D}. In this spacetime particles can be defined in two asymptotically  flat regions. Considering the field in the remote past to be in the vacuum state,
it is found that in the distant future the state contains particles. In the interim region, where the universe is undergoing expansion, no sensible notion of particles exist.  In \cite{JL:Ball:IFuentesSchuller:FPSchuller:06} it was shown that in the future infinity region entanglement has been created between field modes. Interestingly, it is possible to learn about the expansion parameters of the Universe from the entanglement generated. The vacuum state in the past infinity corresponds to the following two mode states in the future infinity
\[\ket{0}^{in}=\sqrt{1-\gamma}\sum_{n}\gamma^n\ket{n}_{k}^{out}\ket{n}_{-k}^{out}.\]
Since the state is pure, we can employ the von Neumann entropy to quantify the entanglement generated the field modes $k$ and $-k$. In order to do this we need to compute the reduced density matrix for one of the modes.
The density matrix for the state is
\begin{eqnarray}
\rho_0&=&\ket{0}^{in}\bra{0}^{in}=(1-\gamma)\sum_{n,m}\gamma^{(n+m)}\ket{n}_{k}^{out}\ket{n}_{-k}^{out}\bra{m}_{k}^{out}\bra{m}_{-k}^{out}.
\end{eqnarray}
The reduced density matrix for mode $k$ is obtained by tracing over mode $-k$
\begin{eqnarray}
\rho_{k}&=&tr_{-k}[\rho_0]=(1-\gamma)\sum_{n,m}\gamma^{(n+m)}\delta_{nm}\ket{n}_{k}^{out}\bra{m}_{k}^{out}\no
&=&(1-\gamma)\sum_{n,m}\gamma^{2n}\ket{n}_{k}^{out}\bra{n}_{k}^{out}.
\end{eqnarray}
Since the reduced denstiy matrix is already in diagonal form with eigenvalues $\lambda_n=(1-\gamma)\gamma^{2n}$, it is straight forward to compute the von Neumann entropy
\begin{equation}
S(\rho_k)=-(1-\gamma)\sum_n\gamma^{2n}\log_2((1-\gamma)\gamma^{2n})
=\log_{2}\left[\frac{\gamma^{\gamma/(\gamma-1)}}{1-\gamma}\right].
\end{equation}
Entanglement has been created in the remote future due to the expansion of the universe. The entanglement depends on the cosmological constants since the coefficient $\gamma$ depends on the expansion rate $\sigma$, the expansion volume $\epsilon$ and the frequency of the modes involved through
\be{gamma}
\gamma=\frac{\sinh^2(\pi\omega_{-}/\sigma)}{\sinh^2(\pi\omega_{+}/\sigma)},
\ee
with
$\omega_{\pm}=1/2(\omega_{out}\pm\omega_{in})$,
$\omega_{in}=[k^2+m^2]^{\frac{1}{2}}$ and
$\omega_{out}=[k^2+m^2(1+2\epsilon)]^{\frac{1}{2}}.$ In the case of  light particles, the equations can be inverted and we can show that we can estimate the expansion parameters from the entanglement.  For a calculation in the covariant matrix formalism see \cite{NFriis:IFuentes}. The analysis of entanglement in an expanding universe has been extended to Dirac fields \cite{PhysRevD.82.045030} and
entanglement in particle creation has also been studied in \cite{Lin2010a}.

\section{Extracting global mode entanglement using point-like detectors}\label{Extracting_global_mode_entanglement}
Although early investigations on relativistic entanglement allowed us to understand some aspects of the observer dependent nature of entanglement,  the results are purely  academic. Global mode entanglement cannot be measured nor manipulated by local observers. Alice would not be able to take an entangled state and apply, for example,  a CNOT gate on it since the states live in the whole spacetime. This would require infinite time and energy making impossible any practical quantum information task. However, it has been shown that global mode entanglement can  be accessed by local observers employing Unruh-Dewitt detectors
\cite{Resnik:03,PhysRevA.71.042104,SYLin:BLHu:10,Olson:2011bq,cqedsabin}. Two Unruh-deWitt detectors at rest can become entangled by interacting with the Minkowski vacuum even if they are space-like separated. This shows that the vacuum state is entangled. The vacuum is not mode-wise entangled however, it contains entanglement in spatial degrees of freedom. This entanglement is extracted, or transferred (swapped in the quantum information language) from the field to the detector's degrees of freedom. In principle, it should be possible to use Unruh-Dewitt detectors to study some aspects of the global mode entanglement analysis within a more practical approach \cite{fuentes}.

Since Unruh-Dewitt detectors are local, one can consider them as suitable systems for the implementation of quantum information tasks. For example, Alice and Bob could be spacelike separated and entangle thier detectors through the interaction with the vacuum. This entanglement could be exploited for quantum information tasks.
The entanglement extracted is currently too small to be useful, but in the future, enhancement schemes may be developed.
At the moment, the main limitation in using Unruh-Dewitt detectors for quantum information processing is that the mathematical techniques involved already become un-manageable  with a few systems. For example, for two detectors one must consider the interaction Hamiltonian
\begin{eqnarray*}
\label{generalhamiltonian2}
\fl \hat{H}_{I}(\tau) &=& \int dk \lambda(\tau_1)(d_1\,e^{i\Omega_1\tau_1}+d_1^\dagger\, e^{-i\Omega_1\tau_1})\left(a_k\,e^{i (k x_1(\tau_1)-\omega t_1(\tau_1))} + a_k^\dagger\,e^{-i (kx_1(\tau_1)-t_1(\tau_1))}\right),\nonumber \\
\fl &+& \int dk \lambda(\tau_2)(d_2\,e^{i\Omega_2\tau_2}+d_2^\dagger\, e^{-i\Omega_2\tau_2})\left(a_k\,e^{i (k x_2(\tau_2)-\omega t_2(\tau_2))} + a_k^\dagger\,e^{-i ( kx_2(\tau_2)-t_2(\tau_2))}\right),\nonumber
\end{eqnarray*}
 for which entanglement calculations involve $4th$-order perturbation theory \cite{PhysRevA.71.042104,SYLin:BLHu:10,Olson:2011bq}. Current research is focused on finding ways to modify the detectors in order to make them more mathematically accessible \cite{Ant}.

 In spite of the mathematical difficulties involved, interesting results have been found in relativistic quantum information using point-like detectors. For example, the entanglement between an Unruh-DeWitt detector and a massless bosonic field has been considered in \cite{Lin2008s} showing that the field and detector remain entangled at late times. Information stored in the point-like system flows into the field  propagating with the radiation into null infinity. The information lost by the detector is never restored (non-Markovian regime).  Another analysis considers two detectors, one inertial belonging to Alice, and one in uniform acceleration belonging to Rob, which do not interact directly but are coupled to the same field \cite{Lin2008v}. The detectors are initially entangled and their entanglement vanishes after a finite time.  The time at which the entanglement vanishes is shorter for higher accelerations, as expected.  This setting has been considered recently for teleportation of a coherent state \cite{Lin2012a}.  When both detectors are at rest and initially entangled, the detectors also disentangle and the dependence with their separation is studied in \cite{Lin2009}. The case of Rob having a detector in non-uniform acceleration has been studied in \cite{Lin2008v}. The authors show that the disentangling time slows down if the motion is non-uniform.  Two uniformly accelerated detectors have also been considered, each on of them in a different Rindler wedge \cite{Resnik:03,SYLin:BLHu:10}. The authors show that entanglement is generated through their interaction with the Minkowski vacuum even if they are always causally disconnected. This shows again that the vacuum contains quantum correlations.
Another interesting analysis is that of the creation and loss of entanglement between detectors moving in circular paths \cite{PhysRevA.81.062320}.

Recently, Olson and Ralph \cite{Olson:2011bq} found that by independently quantizing the past and future spacetime regions of a massless quantum field it is also possible to extract  vacuum entanglement from timelike separated regions. To extract the entanglement they considered two timelike separated point-like detectors at rest characterized by a strongly time-dependent energy gap. Time-like entanglement has been further analyzed in \cite{fixandrzej} and exploited in a teleportation-type protocol \cite{time-liketeleportation,PhysRevLett.109.033602} which can be implemented in a circuit QED experiment \cite{PhysRevLett.109.033602}.

In curved spacetimes the entanglement between point-like detectors can been used to distinguish between two universes. Two inertial detectors in flat spacetime would become entangled if they interacted with a Minkowski thermal field however, they would not become entangled in an exponentially expanding de Sitter spacetime \cite{GSteeg:NCMenicucci:09}.

\section{Accelerated cavities and localized wave-packet  for quantum information}\label{Accelerated_cavities}
Work on the use of Unruh-Dewitt detectors for relativistic quantum information processing has inspired efforts to find suitable systems to store information in the framework of quantum field theory, with the hope of developing mathematical techniques which render the calculations accessible. At the moment moving cavities and localized wave-packets seem plausible systems for this purpose.
\subsection{Entanglement creation and degradation in moving cavities}\label{Entanglement_creation_and_degradation_in_moving_cavities}
The idea of using cavities moving in spacetime for relativistic quantum information processing was proposed in \cite{AlsingMilburn03}  where two observers Alice and Rob share an entangled state of two cavities as a resource for quantum teleportation in non-inertial frames. The mathematical techniques to implement this idea were introduced in \cite{TGDownes:IFUentes:TCRalph:10} for inertial and uniformly accelerated cavities and in \cite{DEBruschi:JLouko:IFuentes:12}  for non-uniform motion.  The authors in  \cite{TGDownes:IFUentes:TCRalph:10} showed that it is possible to entangle two cavities, one inertial and the second in uniform acceleration, by letting a single two level atom interact with the cavities.  The scheme is a generalization of \cite{DEBrowne:MBPlenio:03} in which the modes of two cavities at rest positioned next to each other become maximally entangled after an excited atom moving through them emits an excitation.  Since it is not possible to know in which cavity the atom emitted, the cavities become entangled. When one of the cavities is in uniform acceleration the cavity modes become detuned affecting the atom's emission probabilities.  The modes of an inertial and accelerated cavity are given in (\ref{icmodes}) and (\ref{rcmodes}), respectively.  The interaction Hamiltonian considered by the authors is of  the Unruh-Dewitt type given in (\ref{generalhamiltonian2}) only in the sense that the atom is a two-level ponit-like system interacting with discrete cavity modes. The authors show that the ability to maximally entangle the cavities is reduced with acceleration. However, by appropriately modifying the cavity length it is possible to alter the modes and compensate for the effect.  An advantage of this scheme is that the quantum information stored in cavities is protected from the Unruh effect (unlike for information stored in point-like systems) by the mirrors. As long as the cavity walls are perfectly reflecting the field inside the cavity is independent from the field outside. While the cavity undergoes uniform acceleration (again we consider the case $a L /c^2 \ll 1$ so that the acceleration is essentially constant across the proper length of the cavity), the states within the cavity however, remain pure since the Unruh horizon is not contained in this spacetime region \cite{Schutzhold:Unruh:2005}. The quantum information stored in field modes remains constant at all times as long as the cavities are inertial or in uniform acceleration. However the situation changes when the motion is non-uniform.

Consider a single inertial cavity with all modes in the vacuum state.  If the cavity accelerates for a finite time the modes will become populated as shown is section \ref{Moving_cavities}.
Working in the low acceleration regime $h<1$
it is possible to find analytical expressions for the Bogoliubov coefficients  for bosons \cite{DEBruschi:JLouko:IFuentes:12} and fermions \cite{NFriis:ARLee:DEBruschi:JLouko:10} and calculate the entanglement produced between two modes $k$ and $k^{\prime}$ using the covariant matrix formalism \cite{PhysRevD.85.081701}.  The entanglement quantified by the logarithmic negativity is given by $\mathcal{N}=|\beta^{(1)}_{k,k^{\prime}}|$ which is the first order correction to the Bogoliubov coefficient $\beta_{k,k^{\prime}}$ \cite{NFriis:IFuentes}. The entanglement between the modes can be calculated for any trajectory composed of segments of inertial and uniformly accelerated segments. The entanglement in a basic building block composed of initial and final inertial segments with a single period of uniform acceleration is periodic in the acceleration period.  A consequence of this is that if one maximally entangles two modes each in a different cavity as proposed in  \cite{TGDownes:IFUentes:TCRalph:10} and then accelerates for a finite time, the entanglement between those modes will be degraded since the initial information will be transferred to other modes.  Since the evolution of entanglement is periodic, Rob can plan his travel such that at the end of it his cavity will still be maximally entangled with Alice's.

 Interesting conclusions can be drawn from comparing the analysis of entanglement generation in fermionic and bosonic cavities \cite{PhysRevD.85.081701}. In both cases entanglement is always generated under non-uniform motion however, the entanglement generated is different due to particle statistics. In the bosonic case motion can populate modes with the characteristic that any number of excitations can be produced in a given mode. However, fermionic statistics constrains the number of particles in each mode giving rise to differences in the entanglement produced.

 Another interesting property of motion generated entanglement is that for any trajectory segment constructed from periods of inertial and uniformly accelerated motion, there is an entanglement resonance when the segment is repeated $N$ times \cite{DEBruschi:Dragan:Lee:JLouko:IFuentes:12}. It can be shown analytically that when the frequency associated with the segment period is equal to the sum of the frequencies of the two modes then the entanglement between those two modes grows linearly with $N$. The presence of the resonance is independent of the segment travel details however the degree of entanglement does depend on the specific trajectory. Using the covariant matrix formalisms it is possible to calculate the entanglement at resonance of any trajectory analytically. Interestingly, through these resonances it is possible to produce motion generated quantum gates \cite{DEBruschi:Dragan:Lee:JLouko:IFuentes:12,Friss:Hubber:Bruschi:IFuentes:12,newjorma}. Recent results also show that genuine multipartite entanglement is generated through the periodic motion of the cavity \cite{Friss:Hubber:Bruschi:IFuentes:12}.

The general case of a cavity undergoing an arbitrary acceleration and the possibility of performing actual experiments with mechanically oscillating optical cavities are under study \cite{newjorma}. In the cases of a linear sinusoidal or a uniform circular motion a resonance appears at much lower frequencies for which no new photons are generated. The resonance leads to the generation of entanglement between existing and previously non-entangled cavity modes \cite{newjorma}.

Understanding the effects of gravity on quantum properties and quantum information is of great interest.  The analysis for entanglement in moving cavities already allows us to draw some interesting conclusions: via the equivalence principle \cite{MTW} the results suggest that gravity can create entanglement in the field modes of a cavity (as long as the cavity is small). Consider a cavity in the vacuum held at a fixed position in a uniform gravitational field. The modes of the cavity would become entangled by changing the cavity's position and letting the cavity undergo periods of free fall.

\subsection{Localized wave packets}\label{Localized_wave_packets}
An alternative for constructing localized field states is achieved by superimposing global plane wave modes $U_k(x)=e^{i k_\mu x^\mu}$ localized by a complete set of shape functions $f_m(k)$ via
$U_{m k}(x) = \int dk f_m(k) U_k(x)$ \cite{STakagi:86,Audretsch19944056}. A common example involves ``box normalization" in which one quantizes in a box of length $L$ and imposes periodic boundary conditions, leading to normalized states $f_m^{box} = L^{-1/2} e^{i 2 \pi m x/L}$. One can also normalize in both momentum and space degrees of freedom by using functions \cite{STakagi:86} $f_{m l}(k) = \chi_m(k) e^{-i 2 \pi l k/\epsilon}$
where $\chi_m(k) = \epsilon^{-1/2}$ for $|k-m\epsilon|< \epsilon/2$. Here $m,l$ are integers and $\epsilon$ is a small positive constant with dimension of inverse length. The additional degree of freedom represented by the index $l$ expresses the location of the box. In box normalization one takes the limit as $L\rightarrow\infty$, while in the latter procedure one effectively uses and infinite number of boxes of length $\sim 2\pi\epsilon$.

\subsection{Quantum communication with moving wave-packets}\label{Quantum_communication_with_moving_wave-packets}
As one recently proposed example of a relativistic quantum information protocol, Downes \textit{et al.} \cite{QCommRalph} consider the case of Alice sending coherent states to Rob's accelerated detector, who then
performs homodyne detection. Here the authors bypass the intermediate Unruh modes and consider the full integral Bogoliubov transformation \Eq{Mink2Rindler} from which they form localized modes in frequency,
as in the first equality in \Eq{wavepackets} in the appendix (section \ref{appendix}).
In addition, they eschew the Schrodinger state transformation approach and instead consider
the transformation of observables in the
Heisenberg picture. With Alice stationary at an $x$-coordinate position to the right of focus of Rob's hyperbola
at $x=1/a$ ($\chi=0$), they consider only left moving $L$ modes that travel along geodesics with $x+t$ constant.
Rob's homodyne operator is given by $\hat{O}(\tau) = a_{i,I}^{S}\,a_{i,I}^{L\dag}\,e^{i\phi} + h.c.$ where
the superscripts $K=S,L$ denote the signal and local oscillator modes (both of which are sent to Rob by Alice as localized
Minkowski frequency modes) and the subscript $i$ denotes modes (labeled by components of the wavevector), both parallel (longitudinal) and perpendicular (transverse) to Rob's acceleration. Rob integrates the photocurrent from his detector over a long time compared to the pulse length of
Alice's signal. Thus the average value of the signal received by Rob is given by the expectation value
$X=\langle \int d\tau \hat{O}_i(\tau) \rangle$ with variance $V =\langle (\int d\tau \hat{O}_i(\tau))^2 \rangle - X^2$,
with the Minkowski coherent state $\ket{\alpha,\beta,t}_i = D^S_i(\alpha) D^S_i(\beta)\ket{0}_M$. Here
$D^K_i(\gamma)=\exp(\gamma a^K_i - \gamma^* a^{K\dag}_i )$ is the Minkowski displacement operator with $\gamma = \alpha,\beta$
the amplitude of the signal and local oscillator respectively.  The authors make reasonable approximations for matching the transverse spatial profile of the signal and local oscillator to Rob's detector (``beam-like" communication between Alice and Rob), and further assume that the longitudinal source wavevector is strongly peaked about a central frequency $\om_{so}$. They obtain the result that the normalized variance, defined as the variance $V$ divided by the amplitude of the local oscillator $\beta_R^2$ as seen by Rob is given by the following Planck factor $\bar{V}=V/\beta^2_R = 1/(e^{2\pi\om_{so}(x+t)}-1)$.

The ``novel'' dependence of the Planck factor $\bar{V}$ on $x+t$ is correctly attributed by the authors to the time dependent Doppler shift of the signal frequency $\om_{so}$ sent by Alice as measured by Rob \cite{brout,AlsingMilonni,padmanabhan}. This should not be confused with the usual Planck factor $1/(e^{2\pi\om_{d,R}}-1)$ associated with the spectrum of frequencies $\om_{d,R}$ Rob's detector measures when intercepting a Minkowski plane wave $e^{i\om_{so}(x+t)}$ from Alice.
The two are closely related as follows. The signal sent by Alice is centered on the 4-wavevector $k^\mu_{so} = (\om_{so},-\om_{so},0,0)$, where we have taken $k_{x,so} = -\om_{so}$ for a left moving photon. For Rob's trajectory $x^\mu = 1/a\,(\sinh(a\tau),\cosh(a\tau),0,0)$ his 4-velocity is given by $u^\mu_R = (\cosh(a\tau),\sinh(a\tau),0,0)$. By the previous discussion on Killing vectors in section \ref{Global_fields}, the frequency measured by Rob of this photon is given by $\om_R(\tau) = k^\mu_{so}\,u_{\mu,R} = \om_{so} e^{a\tau}$. By substituting in $x+t$, the variance can be written as  $\bar{V}= 1/(e^{2\pi\om_R(\tau)/a}-1)$, that is, the variance depends on the time dependent Doppler shifted frequency $\om_R(\tau)$ that Rob measures when
he detects a Minkowski plane wave $e^{i\om_{so}(x+t)} = e^{i\om_R(\tau)/a}$
of fixed Minkowski frequency $\om_{so}$.  This Minkowski plane wave is not detected as a positive frequency Rindler plane wave
$e^{-i\om_{d,R}\tau}$ by Rob, but rather as a combination of positive and negative frequency Rindler plane waves
$e^{\mp i\om_{d,R}\tau}$. The square of the Fourier transform $S$ of $e^{i\om_{so}(x+t)}$
as analyzed by Rob $S = \int_{-\infty}^\infty d\tau\,e^{-i\om_{d,R}\tau}\,e^{i\om_R(\tau)/a}$, yields the Planck factor $1/(e^{2\pi\om_{d,R}}-1)$ \cite{AlsingMilonni,padmanabhan}.
In fact, for finite integration times
$S(\tau) = \int_0^\tau d\tau\,e^{-i\om_{d,R}\tau}\,e^{i\om_R(\tau)/a} \equiv \int_0^\tau d\tau\,e^{i\varphi(\tau)}$
the phase of the integrand has a stationary point \cite{brout,AlsingDowlingMilburn05} at $d\varphi(\tau)/d\tau=0$ yielding the condition $\om_{d,R}=\om_{so}\,e^{a\tau}$ which
is the frequency that appears in $\bar{V}$. Previously wave packets for relativistic quantum information where considered in \cite{Bradler2007}. Recently a model on how to detect localized field modes has been introduced \cite{dragan}.
This model can also be used to study entanglement between localized wavepackets in non-inertal frames \cite{dragan2}.


\section{Change of state under Lorentz transformations}\label{LTs}
While the main body of this work has been concerned with observer dependent entanglement for one or more observers undergoing constant acceleration, there is a large body of research on closely related effects for purely inertial observers (the zero acceleration case). This work is loosely classified under the heading of the change in quantum states under Lorentz transformations (LTs), and is intimately tied to the concept of \textit{Wigner rotation} and the unitary transformation of massive and massless particles under the Poincare group (translations, rotations and boosts). In the following, we review this research, which continues to be very active today, especially in connection to the transformation of entangled quantum states in curved spacetime.
\subsubsection{Preliminaries: Wigner rotation}\label{Preliminaries}
In flat (Minkowski) spacetime, the positive energy, single particle states of a massive  particle forms
a spinor representation of the inhomogeneous Lorentz (Poincare) group \cite{weinberg_qft1,tung}
consisting of ten generators: four translations $P^\u = (P^0,P^i)$, three generators $\{J^i\}$ of angular momentum that generate spatial rotations,
and three generators $\{K^i\}$ of pure boosts. The quantum states, denoted by $\ket{\bvec{p},\l}$,
are labeled by their spatial 3-momentum $\bvec{p}$ (where $p^\u = (p^0,\bvec{p})$ with $p^0=E=\sqrt{\bvec{p}^{\sph 2} + m^2}\,$),
and  $\l$ the component of angular momentum along a quantization axis in its rest frame
(typically taken to be along the third ($z$) spatial direction).
Under a Lorentz transformation $\Lambda$ the one-particle state
transforms under the unitary transformation $U(\Lambda)$ as \cite{weinberg_qft1,tung}
\be{1}
U(\Lambda) \ket{\bvec{p},\l} = \sum_{\l'} \,
D^{(j)}_{\l'\l}(W(\Lambda, p))\,\ket{\bvec{p}_\L,\l'},
\ee
where $j$ is total angular momentum of the particle (equal to the spin of the particle in its rest frame),
the summation is over $\l' = (-j,-j+1,\ldots,j)$, and
$\bvec{p}_\L$ are the spatial components of the Lorentz transformed 4-momentum, i.e.
$\bvec{p}^{\pr}$ where $p^{\pr\u} = \Lambda^\u_{\sp\v} \, p^\v$. Most research has been
primarily concerned with massive spin-$\half$  particles $(j=\half)$, and massless spin-1 particles ($j=1$, photons).
While we will primarily discuss spin $\half$ particles, many related issues carry over to the case of photons,
with important technical distinctions due to their absence of mass.
In \Eq{1}, $D^{(j)}_{\l'\l}(W(\Lambda,p))$ is a $(2 j+1)\times (2 j+1)$ matrix
spinor representation of the rotation group $SU(2)$. The quantity $W(\Lambda,p)$ is called the \tit{Wigner rotation} (matrix),
and depends on both the Lorentz transformation $\L$ describing the (constant velocity) inertial reference frame of the observer
and the 4-momentum $p$ of the particle observed.
The important point to note about \Eq{1} is that a single basis state $\ket{\bvec{p},\l}$ is transformed
by a unitary representation of a momentum dependent rotation $W(\Lambda,p)$ into a
superposition of all $\l$ (evaluated at the transformed momentum $\bvec{p}_\L$).

The explicit form of the Wigner rotation matrix is given by
\be{2}
 W(\Lambda,p) =  L^{-1}(\Lambda p) \, \Lambda \, L(p),
\ee
where $L(p)$ is a \tit{standard boost}  taking the standard rest frame 4-momentum $k\equiv (m,0,0,0)$
to an arbitrary 4-momentum $p$, $\Lambda$ is an arbitrary LT taking $p\to \Lambda p$ , and
$L^{-1}(\Lambda p)$ is an inverse standard boost taking the final 4-momentum $\Lambda p$ back to
the particle's rest frame. Because of the form of the standard rest 4-momentum $k$, this final
rest momentum $k^{\pr}$ can at most be a spatial rotation of the initial standard 4-momentum $k$, i.e.
$k^{'} = W(\Lambda,p) \, k$. The rotation group $SO(3)$ is then said to form
(Wigner's) \tit{little group} for massive particles, i.e. the invariance group of the particle's rest 4-momentum.
The explicit form of the standard boost is given by \cite{weinberg_qft1}
\bea{3}
 L^0_{\sph 0}(\xi) &=&  \frac{p^{0}}{m} \equiv \cosh\xi, \no
 L^i_{\sph 0}(\xi) &=& \frac{p^{i}}{m} \equiv \sinh\xi \, \hat{p}^i, \quad L^0_{\sph i} = -\sinh\xi \, \hat{p}_{i}, \no
 L^i_{\sph j}(\xi) &=& \delta^i_{\sp j} - (\cosh\xi-1)\, \hat{p}^i \hat{p}_j, \qquad i,j = (1,2,3),
\eea
where $\gamma = p^{0}/m = E/m$ is the particle's energy per unit rest mass. Note that
for the flat spacetime metric $\eta_{\a\b} =$diag$(1,-1,-1,-1)$, $p_0 = p^0$ and $p_i = -p^i$, and
$\hat{\bvec{p}} = \bvec{p}/|\bvec{p}|=(\sin\theta \cos\phi, \sin\theta\sin\phi,\cos\theta) = \hat{p}^i$
is the direction of the 3-momentum $\bvec{p}$.

In the rest frame of the particle, with (standard) 4-momentum $k=(m,\bvec{k}=\bvec{0})$ the basis vectors $\ket{\bvec{0},\l}$ are eigenstates
of the four momentum $P^\u$, total angular momentum $\bvec{J}^2=j(j+1)$, and third component of angular momentum $J^3$ as
(in units $\hbar=c=1$)
\bea{4}
P^\u \ket{\bvec{0},\l} &=& k^\u \ket{\bvec{0},\l}, \no
\bvec{J}^2\ket{\bvec{0},\l} &=& j(j+1) \ket{\bvec{0},\l}, \no
J^3\ket{\bvec{0},\l} &=& \l \ket{\bvec{0},\l}.
\eea
The standard boost \Eq{3} can be decomposed \cite{weinberg_qft1,tung} as
$L(p)=R(\hbvec{p})\,B_3(|\bvec{p}|)\,R^{-1}(\hbvec{p})$ where $R(\hbvec{p})$ is a rotation
that takes the $\hbvec{z}$-axis to $\hbvec{p}$, and $B_3(|\bvec{p}|)$ is a pure boost in the $\hbvec{z}$ direction such that
$\ket{|\bvec{p}| \hbvec{z},\s}=B_3(|\bvec{p}|)\,\ket{\bvec{0},\s}$ with $|\bvec{p}|=m\,\sinh\xi$.
The unitary representation of $R(\hbvec{p})$ on the Hilbert space of states is given by $U(R(\hbvec{p}))=e^{i\phi J_3}\, e^{i\theta J_2}$.
Since the rightmost rotation in $L(p)$ has no effect on $\ket{\bvec{0},\l}$ we have \cite{tung}
\be{5}
\ket{\bvec{p},\l} = L(p)\,\ket{\bvec{0},\l}=R(\hbvec{p})\,B_3(|\bvec{p}|)\,\ket{\bvec{0},\l} \equiv H(p)\,\ket{\bvec{0},\l}.
\ee
The states $\ket{\bvec{p},\l}$ are eigenstates of
$P^\u$, $\bvec{J}^2$, and the helicity operator $\bvec{J}\cdot\hbvec{P}/|\hbvec{P}|$ with eigenvalue $\l$, the component of the particle's angular momentum along the direction  $\hbvec{p}$ of it's linear momentum \cite{tung},
\bea{6}
P^\u \ket{\bvec{p},\l} &=& p^\u \ket{\bvec{p},\l}, \no
\bvec{J}^2\ket{\bvec{p},\l} &=& j(j+1) \ket{\bvec{p},\l}, \no
\frac{\bvec{J}\cdot\bvec{P}}{|\bvec{P}|} \ket{\bvec{p},\l} &=& \l \ket{\bvec{p},\l}.
\eea
(Note that $J^3\ket{\bvec{p},\l} \neq \l \ket{\bvec{p},\l}$ since $J^3$ now contains a contribution from the orbital motion).
The Wigner rotation in \Eq{2} is given by $W(\Lambda,p) =  H^{-1}(\Lambda p) \, \Lambda \, H(p)$.
The $\ket{\bvec{p},\l}$ are called \tit{helicity} states.
One can also define \tit{spin} states denoted by $\ket{\bvec{p},\s}$  by suitable rotation
$\ket{\bvec{p},\s} = \sum_\l{D^{(j)}_{\l\s}(R^{-1}(\hbvec{p}))\ket{\bvec{p},\l}}$ \cite{caban08}, which are commonly used
in the literature. Here $R^{-1}(\hbvec{p})$ rotates the 3-momentum direction $\hbvec{p}$ to the  $\hbvec{z}$-axis.

The formalism above carries through for massless particles, but with important physical differences.
Since there is no rest frame for a massless particle, the standard 4-momentum is chosen to be $k\equiv (1,0,0,1)$ with
$\bvec{k} = \hbvec{z}$, and basis state $\ket{\bvec{k},\l}$ labeled by $J^3$ as in \Eq{4}
(which hold with substitution $\bvec{0}\to\bvec{k}$ ).
The general state $\ket{\bvec{p},\l}=H(p)\, \ket{\bvec{k},\l}$ is built up in exactly the same fashion as the massive case discussed in \Eq{5}, where now the ``energy" in $B_3(|\bvec{p}|)$ is given by $p^0=|\bvec{p}| = e^{\xi}$. The helicity states $\ket{\bvec{p},\l}$ again satisfy \Eq{6} with $\l$ now restricted to the two values $\pm j$.
The new feature for massless states is that the helicity $\l$ is a Lorentz invariant since only the phase of the state $\ket{\bvec{p},\l}$ changes in a momentum dependent fashion under a LT $\L$
\be{7}
U(\Lambda) \ket{\bvec{p},\l} = e^{-i\,\l\,\theta(\L,\hat{\bf p})}\,\ket{\bvec{p},\l}.
\ee
Here, the notation indicates that $\theta(\L,\hbvec{p})$  depends only upon the LT $\L$ and the direction $\hbvec{p}$ of the particle's momentum, but
\textit{not} upon its frequency $p^0$. The Wigner rotation angle $\theta(\L,\hbvec{p})$ can be computed as $\bra{\bvec{k},\l} W(\L,p) \ket{\bvec{k},\l}$, where again
$W(\Lambda,p) =  H^{-1}(\Lambda p) \, \Lambda \, H(p)$ leaves the standard momentum $\bvec{k}$ invariant
$k^{'} = W(\Lambda,p) \, k$
(Wigner's little group is now $ISO(2)$, rotations in plane perpendicular to $\bvec{k}$). For $j=1$,
$\theta(\L,\hbvec{p})$ describes the amount of rotation experienced by the plane of polarization of a linearly polarized photon
(for example calculations of $\theta(\L,\hbvec{p})$ see \cite{weinberg_qft1,alsing_milburn02,ueda03}).

Returning to the massive case, the spin states $\ket{\bvec{p},\s}$ (and similarly for the helicity states) are normalized as
$\langle{\bvec{p}^{'},\s^{'}}\ket{\bvec{p},\s}= (2\pi)^3\,(2p^0)\delta_{\s^{\pr}\s}\delta^{3}(\bvec{p}^{\pr}-\bvec{p})$.
One-particle states (wavepackets) are given by \cite{tung,PeresTernoRMP}
\be{8}\fl
\ket{\Psi} = \sum_\s \, \int^\infty_{-\infty} d\u(p) \, \psi_\s(p) \, \ket{\bvec{p},\s}, \,\,
\psi_\s(p) = \bra{\bvec{p},\s} \Psi \rangle, \,\,
\sum_\s \, \int^\infty_{-\infty} d\u(p) \, |\psi_\s(p)|^2=1,
\ee
with respect to the Lorentz invariant measure $d\u(p) = d^3\bvec{p}/(2\pi)^3\,(2p^0)$.
Under a LT $\L$, the same state $\ket{\Psi}$ described in the boosted frame is given by
\numparts
\bea{9}
 \ket{\Psi^{'}} &=& \sum_\s \, \int^\infty_{-\infty} d\u(p) \, \psi^{'}_\s(p) \, \ket{\bvec{p},\s}, \\
 \psi^{'}_\s(p) &=&  \sum_{\eta} D^{(1/2)}_{\s\eta}(W(\L,\L^{-1}p)) \, \psi_\s(\L^{-1}p) =
\bra{\bvec{p},\s} U(\L) \ket{\Psi},
\eea
\endnumparts
which can be obtained from a straight forward calculation using \Eq{8} and \Eq{1}, the basis state normalization, a change of variables $p\to \L^{-1}\,p$ and utilizing $d\u(\L^{-1}p)=d\u(p)$, along with some index relabeling.

\subsubsection{Early investigations}\label{Early_investigations}
Quantum information was first considered in a relativistic setting by Czachor in 1997 who analyzed a relativistic version of the Einstein-Podolsky-Rosen-Bohm experiment \cite{alsing_milburn02,ueda03,PeresTernoRMP,czachor,ahn03,massar06,pankovic05,ball06}. Czachor's work showed that relativistic effects are relevant to the experiment where the degree of violation of Bell's inequalities depends on the velocity of the entangled particles. However, it was not until 2002 that the importance of investigating quantum information in the presence of spacetime was recognized. A. Peres \& D. Terno pointed out that most concepts in quantum information theory may require a reassessment \cite{PeresTernoRMP,PeresScudoTerno}. Further interesting results on entanglement in flat spacetime were obtained by
Alsing \& Milburn \cite{alsing_milburn02}, Adami, Bergou, \& Gingrich \cite{adami_spin,adami_photon}, and Solano \& Pachos \cite{pachos03}. Their work showed that although entanglement is conserved overall under a change of inertial frame, it may swap between spin and position degrees of freedom. (For a comprehensive account of this early work up to 2004, with indications of future work on accelerated observers, see the review by A. Peres and D.R. Terno  \cite{PeresTernoRMP}).

The insightful and provoking work of Peres and Terno \cite{PeresTernoRMP,PeresScudoTerno} considered single particle wavepackets as in \Eq{8}  with density matrix $\rho=\ket{\Psi}\bra{\Psi}$ having components $\rho_{\s\s^{'}}(\bvec{p}, \bvec{p}^{'}) = \psi^{*}_{\s}(p) \, \psi_{\s^{'}}(p^{'})$,
and traced out over the momentum to produce the reduced spin density matrix
\be{10}
\rho^{red}_{\s\s^{'}} = \int^\infty_{-\infty} d\u(p) \, \psi^{*}_{\s}(p) \, \psi_{\s^{'}}(p^{'}).
\ee
They showed that owing to the momentum dependent Wigner rotations that arise in the full state \Eq{9}, that are induced by the LT $\L$ describing the reference frame of a new observer, the reduced quantity $\rho^{red}$ has no covariant transformation law, except in the limiting case of sharp momentum. Only the complete density $\rho$ has a covariant transformation law. They concluded that the ``spin state" of a single particle is meaningless if one does not specify its complete state, including momentum variables.
Even though it may be possible to formally define spin in any Lorentz frame, there will be no relationship between the observable expectation values in different Lorentz frames \cite{PeresScudoTerno}.

Alsing and Milburn \cite{alsing_milburn02} and Terashima and Ueda \cite{ueda03} considered an EPR Bell state with factorable pure momentum eigenstates. In a second rest frame of another pair of observers $A,B$ traveling perpendicularly to the initial rest frame, described by the pure boost $\L$, the initial state $\ket{\Psi}$ is described as $\ket{\Psi^{'}}= U_{AB}(\L)\,\ket{\Psi} = U_A(\L) \otimes U_B(\L)\,\ket{\Psi}$. While the state $\ket{\Psi^{'}}$
remains factorable between momentum and spin, the initial Bell state is coherently superposed with other Bell states. The reason is that $U_A(\L)$ and $U_B(\L)$ give Wigner rotation angles of opposite signs for particles traveling in opposite directions. Maximal EPR correlations are recovered if the detectors in the boosted frame are rotated to the new rest frame direction of spin.

Gingrich and Adami \cite{adami_spin} considered wavepackets for two massive spin $1/2$ particles, with similar form and transformation properties $(U_{AB}(\L) = U_A(\L) \otimes U_B(\L))$ as \Eq{8} and \Eq{9}, respectively
\numparts
\bea{11}
\fl \ket{\Psi_{AB}} &=& \sum_{\s\t} \, \int\int d\u(p)\,d\u(q) \, g_{\s\t}(\bvec{p},\bvec{q}) \, \ket{\bvec{p},\s} \, \ket{\bvec{q},\t}, \\
\label{12} \fl g_{\s\t}(\bvec{p},\bvec{q}) &\to&  \sum_{\s^{'}\t^{'}} \, D^{(1/2)}_{\s\s^{'}}(W(\L,\L^{-1}p))
            D^{(1/2)}_{\t\t^{'}}(W(\L,\L^{-1}q))\, g_{\s^{'}\t^{'}}(\L^{-1}\bvec{p},\L^{-1}\bvec{q}).
\eea
\endnumparts
Since LTs entangle the spin and momentum degrees of freedom within a single particle, entanglement can be transferred between them. Their work showed that this is also the case for pairs of particles, and that LTs affects the entanglement between the spins of different particles, as measured by the Wootters concurrence \cite{wootters} on the reduced density matrix formed by tracing out over the momentum of the complete state $\rho_{AB}=\ket{\Psi_{AB}}\bra{\Psi_{AB}}$. Thus, spin-spin entanglement is not a Lorentz invariant.

These works generated an intense study of relativistic EPR correlations for both spin $1/2$  particles(see \cite{PeresTernoRMP,caban10} and references therein) and photons \cite{alsing_milburn02,PeresTernoRMP,adami_photon,caban_remb03} that continues today. Much of the work for spin $1/2$  particles (which we will primarily discuss henceforth) has focused on the relevant covariant observable(s) for spin such that the expectation values obtained from measurements are the same in all inertial frames. Also in question is the meaning and validity of the reduced density matrix, especially in the case of tracing out the momentum from the complete quantum state.
Terno \tit{et al.} \cite{PeresScudoTerno,PeresTernoRMP} argued that in the relativistic setting momentum is a \tit{primary} variable that has relativistic transformation laws that depends \tit{only} on the LT matrix $\L$ that acts on the spacetime coordinates. Quantities such as spin and polarization are \tit{secondary} variables that depend \tit{both} on $\L$ and the primary variable momentum. Thus, even though the reduced density matrix for the secondary variable may be well defined in any coordinate system, it has no transformation law relating values in different Lorentz frames. Further, the unambiguous and seemingly natural construction of a reduced density matrix by means of tracing out over the primary variables is possible only if the secondary variables are unconstrained. In the absence of a general description, a case-by-case treatment is required.

Shortly afterward, Czachor \cite{czachor03}, Caban and Rembielinski \cite{caban_remb05} (see also \cite{saldanha} and references therein) showed that it is possible to define a Lorentz-covariant reduced spin density matrix for single massive particles. Such an object \cite{caban_remb05} contains information about the average polarization of the particle, as well as information about its average kinematical state. For sharp momentum, the reduced density matrix does not change under LTs. However, in the case of an arbitrary momentum distribution \Eq{8} the entropy of the reduced density matrix is in general not a Lorentz invariant. The conclusion of these studies is that while one can define a Lorentz-covariant finite dimensional matrix describing the polarization of a massive particle, in the relativistic case one cannot completely separate out kinematical degrees of freedom. These results ultimately stem from the momentum dependent spin transformations \Eq{1} induced by a LT. Extensions of this line of research have been carried out for states of two spin $1/2$ particles (see e.g. \cite{caban10} and references therein).
Note that as pointed out by Peres and Terno \cite{PeresTernoRMP} an invariant definition of entanglement of a pair of spin $1/2$ particles typically utilizes the center-of-mass ``rest frame" where $\langle\sum\bvec{p}\rangle=0$. However, due to the problem of cluster decomposition \cite{weinberg_qft1} this definition is not adequate for more than two constituent particles since subsets of particles may have different rest frames. This is a difficult and still unresolved problem that has relevance to investigations of relativistic multipartite entanglement for more than two particles.

\subsubsection{Spin observable issues}\label{Spin_observable_issues}
It is worth noting that complications in defining an appropriate reduced spin density matrix arise from the long appreciated non-unique definition of the ``spin" operator for massive particles \cite{terno03,caban_rembl_wlod09,ryder_qft}. Ultimately, this is traced back to the fact that spin is only defined in the rest frame of the particle, where it coincides with the total angular momentum (since for $\bvec{p}=\bvec{0}$ the orbital angular momentum contribution vanishes). For studies of relativistic EPR correlations, most works have settled on some variation of the the spin operator defined from the Paul-Lubanski (PL) vector \cite{ryder_qft} $W_\u = -1/2\epsilon_{\u\v\rho\s}\,J^{\v\rho}\,P^\s$
where $J_{\u\v}$ are generators of the proper orthochronous  Lorentz group with $J_{ij}=-J_{ji} = \epsilon_{ijk}\,J_k$ and
$J_{i0}=-J_{0i} = K_i$, the generators of rotations and boosts, respectively. The two Casimir operators of the Poincare group are given by
$C_1 = P^\u P_\u = m^2$ and $C_2=W^\u W_\u = -m^2\,j(j+1)$ that commute with all the generators $\{P^\u,J^{\u\v}\}$, and label the basis states
$\ket{\bvec{p},\s}$ by their mass $m$ and spin $j$. The PL vector $W^\u = (\bvec{J}\cdot\bvec{P},\,P^0\bvec{J} + \bvec{P}\times\bvec{K})$ transforms as a 4-vector under LTs and is proportional to the total angular momentum in the rest frame of the particle
$W^\u_{rest}=(0,m\bvec{J})$. Hence, spin is well defined in the rest frame of the particle.
In an arbitrary frame, the time component of the PL vector is proportional to the helicity operator $\bvec{J}\cdot\bvec{P}/|\bvec{P}|$.

For observers in an arbitrary inertial frame the spin $\bvec{S}\equiv\bvec{J}-\bvec{L}$ is defined as the difference between the total angular momentum, which is well defined as a generator of the Poincare group, and the orbital angular momentum $\bvec{L}=\bvec{R}\times\bvec{P}$. While the momentum $\bvec{P}$ is a well defined generator of the Poincare group, there is no well defined notion of a position operator (and hence the concept of localizability) in relativistic quantum mechanics. Hence $\bvec{L}$ and $\bvec{S}$ are not uniquely defined. Popular choices for the position operator \cite{caban_rembl_wlod09} include the center-of-mass operator $\bvec{R}_{c.m.}$ and Newton-Wigner position operator $\bvec{R}_{NW}$  defined by
\be{13}
\bvec{R}_{c.m.} = -\half \left[ \frac{1}{P^0}\,\bvec{K} + \bvec{K}\,\frac{1}{P^0} \right], \quad
\bvec{R}_{NW} = \bvec{R}_{c.m.} -  \displ\frac{\bvec{P}\times\bvec{K}}{mP^0(m+P^0)}.
\ee
This leads to the center-of-mass and Newton-Wigner spin operators
\be{14}
\bvec{S}_{c.m.} = \displ\frac{\bvec{W}}{P^0}, \quad
\bvec{S}_{NW} = \frac{1}{m} \left( \bvec{W} - \displ\frac{W^0\bvec{P}}{P^0+m} \right).
\ee
The operator $\bvec{S}_{NW}$ (favored by researchers such as Terno \tit{et al} \cite{PeresScudoTerno,czachor05} and Caban \tit{et al} \cite{caban_remb05,caban10}) satisfies the usual $su(2)$ commutation relations and is the only axial-vector operator that is a linear combination of the PL vector. Using \Eq{13}, a spin observable is defined by $\hbvec{n}\cdot\bvec{S}_{NW}$ as the projection of spin along the unit spatial direction $\hbvec{n}$. The downside of  $\bvec{S}_{NW}$ is that by itself, it is not part of any known 4-vector, nor second rank tensor with well defined transformation properties under LTs. (Note that for both operators $\bvec{S}\cdot\bvec{P} = \bvec{J}\cdot\bvec{P}$ so that we can also write $W^0 = \bvec{S}\cdot\bvec{P}$). The center-of-mass spin operator $\bvec{S}_{c.m.}$ (favored by researchers such as Czachor \cite{czachor,ahn03,PeresScudoTerno,lee04,czachor05,friis10}) does not generate the $su(2)$ spin algebra and its eigenvalues are momentum dependent. Further, in contrast to the operator $\bvec{S}^2_{NW}$, the operator $\bvec{S}^2_{c.m.}$ does not reduce to the second Casimir operator $C_2 = j(j+1)$. Despite these flaws, the spin observable $\hbvec{n}\cdot\bvec{S}_{c.m.}$, has many favorable properties and is often used in the literature, as well as a  normalized (nonlinear) version (introduced by Czachor \cite{czachor}) $\bvec{S}_{CZ}(\hbvec{n})=\hbvec{n}\cdot\bvec{W}/(m^2 + (\hbvec{n}\cdot\bvec{p})^2)^{1/2}$.
A comparison of EPR correlations using both the Newton-Wigner and center-of-mass of spin observables \cite{caban_rembl_wlod09} reveals that for fixed measurement directions, the EPR correlations do not necessarily decrease monotonically with velocity, and may exhibit local extrema for certain configurations (which also holds true for spin 1 particles). Furthermore, this effect occurs for both types of spin observables, and hence appears to be a generic feature of relativistic correlations, and in some cases strongly depends on which definition of spin observable is utilized.

Recent work by Friis \tit{et al}. \cite{friis10} utilized $\bvec{S}_{c.m.}$ for the investigation of a parameterized set of pure states of two spin $1/2$ particles, with various configurations of entanglement between the spin and momentum degrees of freedom. As indicated by earlier works \cite{alsing_milburn02,ueda03,czachor,adami_spin} the maximum EPR correlation can be recovered in any inertial frame if \tit{both} the state and the spin observable are Lorentz transformed.
In an inertial frame in which the particle has 4-momentum $p^\u=(p^0,\bvec{p})$, the authors show the observable defined by
$\hat{n}(p)\equiv (\hbvec{n}\cdot\bvec{S}_{c.m.})/|\l(\hbvec{n}\cdot\bvec{S}_{c.m.})|$, with
$\l(\hbvec{n}\cdot\bvec{S}_{c.m.})$ the eigenvector of the operator $\hbvec{n}\cdot\bvec{S}_{c.m.}$, can also be written as
$\hat{n}_{local}(p) = \hbvec{n}_{local}(p)\cdot\bvec{\s}$. Here $\bvec{\s}$ is the usual 3-vector of Pauli spin matrices, and
$\hbvec{n}_{local}(p) = (\sqrt{1-\b^2}\hbvec{n}_\perp+\hbvec{n}_\parallel)/(1-\b^2(1-\hbvec{n}^2_\parallel))^{1/2}$
can be interpreted as the detector orientation  $\hbvec{n}_{local}$ as seen from the rest frame of the particle (where
$\b=v/c$ is the velocity of the inertial frame).
This later expression can also be written as
$\hat{n}_{local}(p) = [L^{-1}(p) n]^i\,\s_i/|[L^{-1}(p) n]^j|$ where $|[L^{-1}(p) n]^j|$
is the norm of the spatial portion of the Lorentz transformed orientation vector $L^{-1}(p)\,n$.
Here $n=(n^0,\hbvec{n})$ is the 4-vector whose spatial portion $\hbvec{n}$ is the orientation of the detector in
the inertial frame with 4-momentum $p$.
Thus, $L^{-1}(p)$ transforms the orientation 4-vector $n$ back to the rest frame $n_{local}$, whose spatial portion
$\hbvec{n}_{local}$ indicates the local direction of the detector.
Under a general LT $\L$ the orientation 4-vector is given by $n^{\pr\pr}=\L\,n=\L\,L(p) n_{local}$ so that
$\hat{n}^{\pr\pr}(p) = [L^{-1}(\L p) n^{\pr\pr}]^i\,\s_i/|[L^{-1}(\L p) n^{\pr\pr}]^j|=[W(\L,p) n_{local}]^i\,\s_i/|[W(\L,p) n_{local}]^j|$.
Since the Wigner rotation $W(\L,p)$ does not change the norm of $\hbvec{n}_{local}$ we have that the observable $\hat{n}^{\pr\pr}(p)$ transforms under LTs as
$\hat{n}^{\pr\pr}(p)=\big(W(\L,p) \hbvec{n}_{local}\big)\cdot\bvec{\s}=U(\L,p)\,\big(\hbvec{n}_{local}\cdot\bvec{\s}\big)\, U^\dag(\L,p)$.
Thus, the expectation values of observable $\hat{n}^{\pr\pr}(p)$ are the same as that
in the rest frame of the particle $\hat{n}_{local}(p)$.
Most importantly, while the spin observable $\hat{n}^{\pr\pr}(p)$ depends on the momentum of the particle, the measurement direction $\hbvec{n}^{\pr\pr}$ corresponding to this observable does not \cite{friis10}.

Recently Saldanha and Vedral \cite{saldanha} have raised concerns about the physical implementations of the above spin observable measurements. They argue that while the above formulations are mathematically covariant, in the sense that spin measurements obtain values that are the same in all inertial frames, such measurements may be physically inconsistent. Consider a covariant description of the interaction $H_{PL}$ of a measurement apparatus with spin observable $\bvec{S}$ constructed from the PL vector $W^\u = (W^0,\bvec{W})$, where from \Eq{14} $W^\u = (\bvec{S}_{c.m.}\cdot\bvec{P}, P^0 \bvec{S}_{c.m.})$ or
$W^\u = (\bvec{S}_{NW}\cdot\bvec{P}, m \bvec{S}_{NW} + (P^0-m)(\bvec{S}_{NW}\cdot\bvec{P})\bvec{P}/|\bvec{P}|^2 )$.
The interaction must be a Lorentz scalar of the form $H_{PL} = W^\u\,G_\u = W^0\,G^0 - \bvec{W}\cdot\bvec{G}$ for some 4-vector $G^\u = (G^0,\bvec{G})$ in order that the expectation values of a spin measurement are the same in all inertial reference frames. The authors state there are no known couplings of this type, and in particular, not for measurements where spin couples to the electromagnetic field. They argue, as have other researchers discussed above, that it is not possible to uniquely measure the spin of a particle independent from its momentum. They further conclude that any spin-momentum partition by means of reduced density matrices is ``actually completely meaningless," while other researchers have argued that such bipartite partitions must be examined on a case by case basis, where meaningful information can be obtained.

\subsubsection{Extension to curved spacetime}\label{Extension_to_curved_spacetime}
The extension of the above work to curved spacetime for both spin $1/2$ particles and photons has been examined by Terashima and Ueda \cite{ueda04}, Alsing \tit{et al.} \cite{alsing_WigRot_spinhalf,alsing_WigRot_photon}, Brodutch and Terno \cite{terno11} and Palmer \tit{et al.} \cite{palmer}. In flat spacetime the inertial frame, described by the the LT $\L$ is global. In curved spacetime,
this is true only locally at each spacetime point $x = (x^0,\bvec{x})$, where by the Equivalence Principle \cite{carroll,MTW}, spacetime is locally (Lorentzian) flat, and the rules of special relativity hold.
The LT is now parameterized by the spacetime location $x$ as $\L^\u_{\sp\v}(x)$ - a \textit{local Lorentz transformation} (LLT) which describes transformations between the instantaneous states of motion of different local observers at the same spacetime point $x$, for example freely falling, stationary (e.g. fixed radial coordinate), circular motion, or any trajectory with arbitrary acceleration. These LLTs are to be distinguished from  general coordinate or \tit{world transformations}  $\partial x^{\pr\u}/\partial x^\v$ which relates vectors, tensors, etc... describing the same spacetime quantity in different coordinates $x$ and $x^\pr$. The local Lorentz frame (LLF), i.e. the observer's \textit{local laboratory} \cite{JBHartle:GravBook}, has a small finite extent in space and time in as much as the curvature can be considered constant within this region. The LLF can be described by a \tit{tetrad} $e^{\hat{a}}_{\sp\a}(x)$ \cite{MTW,Wienberg:GravBook,JBHartle:GravBook},
a set of 4 (orthonormal) local axes $\{e^{\hat{a}}(x)\}$ labeled by (hatted local Lorentz) indices $(\hat{0},\hat{1},\hat{2},\hat{3})$ with spacetime components (world coordinate indices) $\a = (0,1,2,,3)$. The spatial triad $\{e^{\hat{i}}(x)\}$ describes the orientation of the LLF in the surrounding spacetime, and the temporal axis ${e^{\hat{0}}(x)}$, equal to the observer's (world) 4-velocity (and thus tangent to their worldline), describes the observer's local proper time. As the observer moves through spacetime their LLF, described by their tetrad, twists and turns in the surrounding curved background. The observer describes events in their local laboratory by projecting world tensors onto this tetrad, e.g. a particle with momentum $p^\a(x)$ passing through the observer's local laboratory is described locally as $p^{\hat{a}}(x) = e^{\hat{a}}_{\sp\a}(x)\,p^\a(x)$.

The quantum mechanical single particle states are now represented as $\ket{p^{\hat{i}}(x),\s}$ where $p^{\hat{i}}(x)$ are the spatial components of the particle's 4-momentum
$p(x) = p^{\hat{a}}(x)\,e_{\hat{a}}(x)$ (also equal to $p^\u(x)\,e_\u(x)$ where $e_\u(x)$ can be taken as the coordinate basis vectors in the surrounding spacetime $e_\u(x)=\partial_{x^\u}$).
Terashima and Ueda  \cite{ueda04} (see also appendix in Alsing \tit{et al.} \cite{alsing_WigRot_spinhalf}) showed that as the observer
with 4-momentum $p(x)$ at $x$ travels to a new, infinitesimally close spacetime point $x^\pr$ in proper time $d\t$, the change $\d p(x) = \d p^{\hat{a}}(x)\,e_{\hat{a}}(x)$ relative to the LLF can be written as a LLT
$\d p^{\hat{a}}(x) =  \l^{\hat{a}}_{\sp\hat{b}}(x) \,  p^{\hat{b}}(x)\,\d\t$, so that $\L^{\hat{a}}_{\sp\hat{b}}(x) = \d^{\hat{a}}_{\sp\hat{b}} + \l^{\hat{a}}_{\sph\hat{b}}(x)\, d\t$.
One can now calculate the effect of the Wigner rotation expressed as a local version of \Eq{1} (with $\bvec{p}\to p^{\hat{i}}(x)$, $\L\to\L(x)$, etc...) and \Eq{2} with
an infinitesimal Wigner rotation given by
$W^{\hat{a}}_{\sp\hat{b}}(x) = \d^{\hat{a}}_{\sp\hat{b}} + \vartheta^{\hat{a}}_{\sp\hat{b}}(x)\, d\t$, where the antisymmetric $\vartheta^{\hat{a}}_{\sp\hat{b}}(x)$ only has non-zero components on the spatial indices (i.e. a rotation). Finite Wigner rotations are obtained by a time-ordered integration along the observer's worldline with respect to his proper time.
The implications of these results is that the orientation of the spin of the particle in the observer's LLF depends on instantaneous state of his motion as embodied in his tetrad $e^{\hat{a}}_{\sp\a}(x)$ describing his local laboratory. This is not unexpected from the results of flat spacetime investigation when one invokes the Equivalence Principle.
However, the entanglement of a pair of spin $1/2$ particles (initially created at a single spacetime point) when the parties are widely spacelike separated, depends not only upon the initial state of entanglement amongst the spin and momentum degrees of freedom (as in the flat spacetime case), but also upon the full history of the motion through spacetime.
That is, Wigner rotation induced spin-momentum entanglement occurs at each spacetime point along the trajectories of the constituent particles. While general relativity advocates an agnostic description of physics using arbitrary observers, the work of Alsing \tit{et al.} \cite{alsing_WigRot_spinhalf} and Palmer \tit{et al.} \cite{palmer} shows that in the \tit{Fermi-Walker frame} (the instantaneous, co-moving, non-rotating rest frame of the observer \cite{MTW}) the Wigner rotation due to the effects of the gravitation field is zero. Though it appears contrary to the spirit of general relativity to single out a particular reference frame as special, the Fermi-Walker frame may prove advantageous to describe internal quantum mechanical degrees of freedom. While in the above discussion we have primarily concentrated on the Wigner rotation for massive particle in curved spacetime, many interesting results follow from the consideration of photons in an arbitrary gravitational field \cite{alsing_WigRot_photon,terno11,palmer,caban07,bradler09}.

\section{Open questions and future directions}\label{Open_questions}

In prototypical quantum communication protocols Alice and Bob produce entangled resources (for example two entangled qubits) to be used to transit information.  Alice and Bob take their systems to their labs. In case spacetime is flat and Alice and Bob move at non-relativistic speeds the entanglement resources they produced will remain un-changed.  However, what happens to quantum resources if Alice and Bob live in curved spacetime? If they move in the presence of a gravitational field? If they move at relativistic speeds? These are some of the most interesting questions in the field of relativistic quantum information.

Understanding quantum information in relativistic settings might lead to finding ways to exploit relativity to improve quantum information tasks. A first step has been taken,
finding suitable systems to store and process information in quantum field theory. The systems presented in this paper seam to be suitable candidates for this and hopefully ideas presented here will stimulate other researchers to find new systems and develop mathematical techniques that will allow us to find answers to the questions posed above. Quantum protocols such as teleportation are possible thanks to the tensor product structure of the Hilbert space. Constructing new protocols which are only possible in a quantum and relativistic world would indeed be exciting.

An interesting question without doubt is if a covariant notion of entanglement can be constructed.
The analysis of entanglement in relativistic settings has been performed by using the standard non-relativistic definition of entanglement and applying it to relativistic settings. By doing this we have found the entanglement is observer-dependent and that inconsistencies my arise when defining subsystems in special relativity. An open question is if there exits a more general notion of quantum correlations in relativistic quantum theory which coincides with the standard notion in the non-relativistic regime.
The answer to this question is intimately tied to issues related to
the physical implementation of local measurements for observers in arbitrary motion.
In this respect, the current research on a locally covariant formulation of quantum field theory in curved spacetime may prove useful \cite{Fredenhagen1102.2376,Fewster1105.6202,TGDownesThesis2011}.

Finally, we would like to say that we find the time ripe to find experimental demonstration of results in the field of relativistic quantum information. On one hand, cutting edge experiments in quantum information are approaching regimes where relativity starts playing a role and on the other hand, recent progress in the experimental demonstration of the dynamical Casimir effect \cite{CMWilson:etal:11,DFaccio:ICarusotto:11} might provide the requisite techniques to show that gravity has effects on entanglement .

\subsection{Acknowledgments}
IF would like to acknowledge the support by EPSRC [CAF Grant EP/G00496X/2].
PMA would like to acknowledge the support of the Air Force Office of Scientific Research (AFOSR) for this work.
We thank A. Dragan, J. Louko, D. Bruschi, N. Friis. A. Lee,  E. Martin-Martinez, G. Adesso, J. Doukas, T.G. Downes, T. Ralph, R. Mann, F. P. Schuller, C. Sabin, R. Jauregui, P. Barberis-Blostein and D. Sudarsky for stimulating conversations.
We strived to be inclusive in our citations to this  exciting and burgeoning field. Any omissions on the part of the authors are purely unintentional.
Any opinions, findings and conclusions or recommendations
expressed in this material are those of the author(s) and
do not necessarily reflect the views of AFRL.

\section{Appendix}\label{appendix} A brief discussion of some of the states used in the literature to investigate
acceleration/observer dependent entanglement is in order.
A highly idealized scenario is to invoke the ``single-mode approximation," (SMA) as considered
by Alsing and Milburn \cite{AlsingMilburn03}, in which Minkowski annihilation operator  is taken to be one of the
right or left moving Unruh modes (since they do not couple to each other) in the integrand of (\ref{Mink2Unruh})
\be{SMA}
a_{\omega,M} \sim a_{\Omega,R} = \cosh(r_\Omega) a_{\Omega,I} - \sinh(r_\Omega) a^\dag_{\Omega,II}.
\ee
One can then consider the entangled Minkowski state between Alice and Bob (left and right kets, respectively)
\be{MinkBellState}
\ket{\Phi} = \frac{1}{\sqrt{2}}\,
\big(
\ket{0_\om}^\M\,\ket{0_{\om^{\prime}}}^\M + \ket{1_\om}^\M\,\ket{1_{\om^{\prime}}}^\M
\big)
\ee
when undergoes uniform acceleration (transforming into his Rindler alter-ego Rob) in wedge $I$.
One can then expand Bob's Minkowski states $\ket{0_{\om^{\prime}}}^\M$ and $\ket{1_{\om^{\prime}}}^\M=a^\dag_{\omega^{\prime},M}\ket{0_{\om^{\prime}}}^\M$
in terms of the Minkowski vacuum expressed in terms of the Rindler wedge modes $I$ and $II$ given above
and the adjoint of (\ref{SMA}). This was done for idealized leaky cavities by Alsing and Milburn\cite{AlsingMilburn03} and for
free space modes by Fuentes and Mann \cite{Alice} (see also the original Alsing and Milburn arxiv version in \cite{AlsingMilburn03}).

Although the SMA gives degradation of entanglement with increasing acceleration, it is not rigourously
justifiable since the Bogoliubov phase factors $\alpha_{\om\Om}^R \sim  (\om l)^{i\eps\Om} = e^{i\eps\Om \ln(\om l)}$
are not localized in frequency (the SMA was invoked to qualitatively represent the central frequency of some wavepacket).
A more refined approximation was considered by Bruschi \textit{et al.} \cite{BSMA} by considering the pre-accelerated
initial Minkowski state to be composed of Unruh modes for Bob (compare with \Eq{MinkBellState})
\be{BSMAState}
\ket{\Phi'} = \frac{1}{\sqrt{2}}\, ( \ket{0_\om}^\M\,\ket{0_\Om}^\U + \ket{1_\om}^\M\,\ket{1_\Om}^\U ).
\ee
The authors considered essentially a generalized version of the integrand in \Eq{Mink2Unruh} by taking
\be{BSMAa}
a_{\omega,M} = q_R \, a_{\Omega,R} + q_L a_{\Omega,L}, \nonumber
\ee
with $|q_R|^2+|q_L|^2=1$, which reduces to the SMA \Eq{SMA} for $q_L=0$. Again, entanglement was degraded as the acceleration
is increased.

The authors further considered the more physically realistic case of states with a finite spread of frequencies, i.e. wavepackets (see \cite{STakagi:86}), versus the highly idealized, sharply defined frequency states considered above (the latter of which capture the essence of the acceleration-dependent entanglement). Instead of \Eq{Mink2Unruh} one considers
\be{wavepackets}\fl
a^\dag_M = \int_0^\infty d\om \, f(\om)\, a^\dag_{\omega,M} =
\int^\infty_0 d\Omega\, [ g_R(\Om)\, \alpha_{\omega\Omega}^R \, a^\dag_{\Omega,R} +
                          g_L(\Om)\, \alpha_{\omega\Omega}^L \, a^\dag_{\Omega,L}] = a^\dag_R +  a^\dag_L,
\ee
where $g_R(\Om) = \int_0^\infty d\om \, f(\om)\,\alpha_{\omega\Omega}^R$ and
$g_L(\Om) = \int_0^\infty d\om \, f(\om)\,\alpha_{\omega\Omega}^L$. By considering  shape functions
that are Gaussian in $ln(\om l)$ and adjusting parameters to ensure that the wavepacket version of the state \Eq{BSMAState} has
negligible overlap between Alice's and Rob's state during acceleration, the SMA  \Eq{SMA} is recovered with one of the $q_R$, $q_L$ vanishing (see \cite{BSMA} for further details).

The two global initial states $\ket{\Phi}$ \Eq{MinkBellState} and $\ket{\Phi'}$ \Eq{BSMAState} consider different scenarios when the second party Bob undergoes uniform acceleration (to become the accelerated Rob).  Both states $\ket{\Phi}$ and
$\ket{\Phi'}$ are maximally entangled, global Minkowski states.
The first state $\ket{\Phi}$ is a ``fixed" Minkowski state, independent of the acceleration $a$ for both Alice and Bob, and is composed of the usual Minkowski plane wave $u_{\om,M}\sim e^{-i\omega(t-\epsilon x)}$ in \Eq{eq:planem}.
The second state $\ket{\Phi'}$ is composed of Minkowski states for Alice, and Unruh states for Bob whose mode $u_{\Om,U}$ is a
linear combination of Rindler modes $u_{\Om,I}$ and $u^{*}_{\Om,II}$ that are both proportional to $\Om^{-1/2} (x-\eps t)^{i\eps\Om}$ in region $I$ and $II$, respectively. On the one hand, by the transformation between Minkowski and Unruh modes \Eq{Mink2Unruh}, $\ket{\Phi'}$ Bob's portion of the state can be considered as an involved integration of global Minkowski modes. The entanglement in this state is then explored  between the inertial Alice, and the Rindler observer Rob, who is specified by a particular value of the acceleration $a$, and whose mode function in region $I$ is $u_{\Om,I}$.
On the other hand, as discussed in section \ref{Flat_spacetime_entanglement} on flat spacetime entanglement, the state $\ket{\Phi'}$
may also be considered as a family of maximally entangled states parameterized by the dimensionless variable $\Omega=\omega/a$ in non-inertial frames. By fixing the physical frequency $\omega$ and changing $a$ one analyzes the entanglement in a family of states which, all have the same frequency $\omega$ as seen by observers with different proper acceleration $a$.  Alternatively, the entanglement corresponds to a family of states with different physical frequency $\omega$ as seen by the same observer moving with fixed proper acceleration $a$ \cite{PhysRevD.86.025026}.

\section*{References}
\bibliographystyle{iopart-num}
\bibliography{ObsDepEntproofs}
\end{document}